\documentclass{article}
\usepackage{eurosym}
\usepackage{amsfonts}
\usepackage{amssymb}
\usepackage{graphicx}
\usepackage{amsmath}
\usepackage{ulem}
\usepackage{subcaption}
\usepackage{color}

\newtheorem{theorem}{Theorem} [section]

\newtheorem{example}[theorem]{Example}

\newtheorem{lemma}[theorem]{Lemma}

\newtheorem{proposition}[theorem]{Proposition}

\newenvironment{proof}[1][Proof]{\textbf{#1.} }{\ \rule{0.5em}{0.5em}}
\allowdisplaybreaks[1]

\begin{document}

\author{N. Allouch\thanks{%
University of Kent - School of Economics, E-mail: N.Allouch@kent.ac.uk},
Luis A. Guardiola\thanks{%
Universidad de Alicante. Departamento de Matemáticas, E-mail: luis.guardiola@ua.es} $^{,} $\thanks{%
Corresponding author.} \ and A. Meca\thanks{%
Universidad Miguel Hern\'{a}ndez de Elche-I.U. Centro de Investigaci\'{o}n
Operativa, E-mail: ana.meca@umh.es}}
\title{Measuring productivity in networks: A game-theoretic approach}
\maketitle

\begin{abstract}

Measuring individual productivity (or equivalently distributing the overall productivity) in a network structure of workers displaying peer effects has been a subject of ongoing interest in many areas ranging from academia to industry. In this paper, we propose a novel approach based on cooperative game theory that takes into account the peer effects of worker productivity represented by a complete bipartite network of interactions. More specifically, we construct a series of cooperative games where the characteristic function of each coalition of workers is equal to the sum of each worker intrinsic productivity as well as the productivity of other workers within a distance discounted by an attenuation factor. We show that these (truncated) games are balanced and converge to a balanced game when the distance of influence grows large. We then provide an explicit formula for the Shapley value and propose an alternative coalitionally stable distribution of productivity which is computationally much more tractable than the Shapley value. Lastly, we characterize this alternative  distribution based on three sensible properties of a logistic network. This analysis enhances our understanding of game-theoretic analysis within logistics networks, offering valuable insights into the peer effects' impact when assessing the overall productivity and its distribution among workers.


\end{abstract}

\newpage

\section{Introduction}

Game theory and network productivity are two fields that have been applied
to the study of logistics networks. In general, game theory is a branch of
mathematics that studies  strategic decision making in various  interactions, while network
productivity is concerned with the efficiency and effectiveness of networks.
In the context of logistics, these fields have been used to study how
decisions made by individual actors within a supply chain can affect the
overall efficiency and productivity of the network. Some possible advances
in this area could include the development of new mathematical models and/or
algorithms to analyze logistics networks, the application of game theory and
network productivity principles to real-world logistics problems, or the
integration of these fields with other areas of logistics research.


This paper focuses on analyzing the measurement of worker productivity in a
logistics network represented by a complete bipartite network. Such a
network structure is particularly interesting  from the perspective of cooperative game
theory, as  all   its induced sub-networks   maintain   the  same  network structure.  This network structure  effectively simulates various productive
and logistical relationships, such as the interconnection between goods
suppliers and consumers, where collaboration between both groups is crucial
for efficient provision of goods. Another real-world example of a complete
bipartite network could be a food supply system connecting producers with
retailers, with producers forming one team and retailers forming the other.
Additionally, the concept is applicable to the internal structure of
companies, where work teams are divided into two fully connected groups. In
this scenario, the network's efficiency depends not only on the individual
productivity of workers within each team but also on the connectivity and
collaboration between the two teams.

Measuring the productivity of workers in a network is crucial as it enables
the identification (and reward) of the most effective employees in their roles. This
knowledge empowers managers to concentrate their resources and training
initiatives on those individuals who require performance improvement.
Additionally, productivity measurement assists in detecting bottlenecks
within the network and areas where efficiency enhancements can be made. Such
insights aid managers in making informed decisions regarding network
re-organization or job reassignments to enhance overall efficiency.

The  network productivity can be thought of   as a public good   for  several reasons. Firstly, network productivity is essential for the efficient functioning and provision of common goods across various contexts such as the environment, health, and logistics. Given that common goods are accessible to the majority of society, network productivity is crucial to ensuring the availability and effective distribution of these goods. Secondly, network productivity is based on the interconnection and collaboration among different actors, including both public institutions and private companies. In the context of providing common goods, actors must work together and share resources to achieve optimal results. Network productivity plays a fundamental role in optimizing interactions and collaboration among these actors, contributing to the efficient provision of common goods.

Moreover, an increase in network productivity can generate positive externalities that benefit society as a whole. For example, higher productivity in logistics can lead to more efficient delivery of common goods, such as medical supplies during a health crisis. These positive externalities have a beneficial impact on society by improving quality of life and contributing to economic and social development. Lastly, effective provision of common goods often requires collaboration between the public and private sectors. Network productivity is a critical component in facilitating cooperation and synergy between these actors, enabling them to coordinate efforts, share resources, and optimize the provision of common goods, especially in crisis or emergency situations. Finally, each agent  intrinsic  productivity can be viewed as public good that   provide different (based on network position), non-rival benefits to all members of society.


    All the above examples share a common theme: measuring the
productivity of agents within a network can help identify opportunities to target interventions in  the network. 
The paper \cite{S09} addresses a gap in the current literature on communication networks
by presenting a unique tutorial on the application of cooperative  game
theory. It comprehensively covers the theory and  technical aspects, and
provides practical examples drawn from game theory and communication applications. Within \cite{Z16}, a cooperative game theory-driven method is proposed, specifically focusing on community detection in social networks. Individuals are viewed as
players, and communities are seen as coalitions formed by players. The
authors use a utility function to measure preference and propose an
algorithm to identify a coalition profile with maximal utility values.
Experimental results demonstrate the effectiveness of the approach.
In \cite{P21}, the authors investigate a cooperative differential game model applied to networks, where players have the ability to sever connections with their neighboring nodes. This enables the evaluation of a characteristic function that measures the value
of coalitions based on cooperation. The authors prove the convexity of the
game, ensuring  the Shapley value belongs   to the  core.

In this paper, we explore a cooperative game framework that considers the influence of peer effects on worker productivity in complete bipartite networks. The investigation into peer effects has recently undergone expansion within networks (refer to \cite{BDF20} for a recent survey). Our analysis focuses on a series of cooperative games where each worker's characteristic function incorporates their own productivity and the productivity of nearby workers within a specified distance. The interconnections are weighted using an attenuation factor, highlighting the impact of neighboring workers on an individual's overall productivity. We show that these games are  balanced  and converge  to a balanced game when the distance of influence grows large  provided that  the attenuation factor is  below  a certain threshold.
 
  We propose three different approaches to distributing productivity among workers. The first approach is   the status  quo    granting each work his  individual  productivity,  which accounts  for peer effects. The second approach utilizes the Shapley value  to   share the overall  productivity, while the third approach, called the Link Ratio Productivity Distribution (LRP distribution), takes
into account the network's structure and the connectivity of the workers. We
characterize the LRP distribution and analyze its impact on the efficiency
of the logistics network. Our study emphasizes the significance
of measuring productivity  of   workers  in a logistics network    represented by a   complete
bipartite network and explores how to distribute the   overall   productivity to
individual  according to their contributions. This analysis contributes to enhancing our understanding of game-theoretic networks within logistics systems, offering insights into the peer effects' impact when assessing the overall productivity and its distribution among workers.

The utilization of cooperative games based on network elements to establish
objective criteria for benefit/cost sharing among network members is a
well-established topic in the literature. In \cite{DP94}, authors
examine different solution concepts in cooperative game theory using a
graph-based game, demonstrating the computational complexity of core
computation and the potential undecidability of the existence of von
Neumann-Morgenstern solutions. The proposed approach in the study by \cite{J05} introduces allocation rules for network games that consider possible changes in the network structure made by players. These rules allocate value based on alternative network structures, providing a comprehensive analysis of the dynamics within network games. The research conducted by \cite{S05} analyzes reward games in network structures, investigating link
monotonic allocation schemes and characterizing conditions for link
monotonicity in the Myerson and position allocation schemes. In the work by \cite{H08}, the average tree solution is presented as a unique solution for cooperative games with communication structures depicted by undirected graphs. The study demonstrates that the game possesses a non-empty core, and under the concept of link-convexity (a weaker condition than convexity), the average tree solution resides within the core. This research provides valuable insights into the solvability and stability of cooperative games within communication networks. The authors in \cite{BJ16} propose algorithms that detect and eliminate the most influential node in order to weaken leadership positions. They employ a greedy approach based on modifying the network's structure. To measure a node's leadership, they utilize the Shapley value and develop algorithms for overthrowing leaders. For   further   information, we recommend
consulting the surveys by \cite{B14,B19}.

The structure of the paper is as follows. It begins with a preliminary
section introducing cooperative game theory and networks. Section 3
describes finite attenuation network games (FAN games) and examines
their main properties. In Section 4, the focus is on establishing a
necessary and sufficient condition for FAN games to converge to a new class
of cooperative games: attenuation network games (AN games), which are shown
to be totally balanced and convex. A coalitionally stable productivity
sharing distribution based on network-generated productivity is also
presented, along with an explicit form of the Shapley value in relation to
the network structure. Section 5 explores an alternative productivity
distribution that considers network structure and worker connectivity,
providing an easier calculation method than the Shapley value. The concept
of difference games, obtained by subtracting consecutive FAN games, is
introduced, and the analysis demonstrates how productivity increases with
distance. A series of distributions for the difference games is  proposed,
converging to an overall productivity distribution for AN games known as the link
ratio productivity distribution (LRP distribution). The coalitional
stability of LRP is established, and it is characterized based on three
desirable properties for a realistic and functional network. Finally,
Section 7 discusses implications and suggests potential avenues for future
research in the field, catering to both academics and practitioners.

\section{Preliminaries}

\bigskip To ensure clarity, we have incorporated in this section the
fundamental principles of cooperative game theory and graph theory that are
essential for comprehending and validating the findings presented in the
paper.

A cooperative (profit) TU-game is a pair $(N,v)$ where $N=\{1,2,...,n\}$ is
a finite set of players. The set of all coalitions $S$ in $N$ is represented
by $\mathcal{P}(N)$, and the characteristic function $v:\mathcal{P}%
(N)\longrightarrow 
\mathbb{R}
$ is defined such that $v(\emptyset )=0$. The value $v(S)$ denotes the
maximum profit obtainable by coalition $S\subseteq N$, where $N$ is commonly
referred to as the grand coalition. The profit vector or allocation is
denoted as $x\in 
\mathbb{R}
^{\left\vert N\right\vert }$, where $\left\vert N\right\vert $ refers to the
cardinality of the grand coalition. We also denote $s=|S|$ for simplicity.

A TU-game $(N,v)$ is considered monotone increasing if larger coalitions
receive more significant benefits, which is expressed as $v(S)\leq v(T)$ for
all coalitions $S\subseteq T\subseteq N.$ Additionally, the game is said to
be superadditive if the benefit obtained by the combination of any two
disjoint coalitions is at least as much as the sum of their individual
benefits. Specifically, $v(S\cup T)\geq v(S)+v(T)$ holds for all disjoint
coalitions $S,T\subseteq N$. It is noteworthy that in superadditive games,
it is reasonable for the grand coalition to form. This is because the
benefit acquired by the grand coalition is at least as great as the sum of
the benefits of any other coalition and its complement, i.e., $v(N)\geq
v(S)+v(N\setminus S),$ for all $S\subseteq N.$

The set of all vectors that efficiently allocate the benefits of the grand
coalition and are coalitionally stable is referred to as the core of the
game $(N,v)$, which is denoted as $Core(N,v)$. More specifically, no player
in the grand coalition has an incentive to leave, and each coalition is
guaranteed to receive at least the profit allocated by the characteristic
function: 
\begin{equation*}
Core(N,v)=\left\{ x\in \mathbb{R}^{\left\vert N\right\vert }:\sum_{i\in N}x_{i}=v(N)\text{
and }\sum_{i\in S}x_{i}\geq v(S)\ \text{\ for all }S\subset N\right\} .
\end{equation*}

\bigskip A TU-game is classified as balanced only when the core is nonempty,
as detailed in \cite{B63} and \cite{SH67}. If the core of every subgame is
nonempty, the game $(N,v)$ is considered to be a totally balanced game (see \cite%
{SS69}). A game $(N,v)$ is regarded as convex if for all $i\in N$ and all $%
S,T\subseteq N$ such that $S\subseteq T\subset N$ with $i\in S,$ then $%
v(S)-v(S\setminus \{i\})\geq v(T)-v(T\setminus \{i\}).$ It is widely
acknowledged that convex games are superadditive, and superadditive games
are totally balanced. Shapley establishes in\cite{SH71} that the core of
convex games is  large enough.

A single-valued solution $\varphi $ is an application that assigns to each
TU game $(N,v)$ an allocation of $v(N)$, the profit obtained by the grand
coalition. Formally, $\varphi $ is defined as follows: $\varphi
:G^{N}\longrightarrow \mathbb{R}^{\left\vert N\right\vert }$, where $G^{N}$
is the set of all TU-games with player set $N$, and $\varphi _{i}(v)$
represents the profit assigned to player $i\in N$ in the game $v\in G^{N}$.
Hence, $\varphi (v)=(\varphi _{i}(v))_{i\in N}$ is a profit vector or
allocation of $v(N)$. For a comprehensive understanding of cooperative game
theory, we recommend referring to \cite{G10}.

The Shapley value, first introduced in \cite{SH53}, is a widely recognized
single-valued solution in cooperative game theory. The Shapley value of
convex games always belongs to the core and it is the baricenter of the core
(see \cite{SH71}). Moreover, it is a linear operator on the set of all TU games.
For a profit game $(N,v)$, $\phi $ is defined as $\phi (N,v)=(\phi
_{i}(N,v))_{i\in N}$, where for each $i\in N$

\begin{equation*}
\phi _{i}(N,v)=\sum\limits_{S\subseteq N\backslash \{i\}}\frac{s!(n-s-1)!}{n!%
}\cdot \left[ v(S)-v(S\setminus \{i\})\right] .
\end{equation*}

We consider a network \textbf{g} of $N=\{1,2,...,n\}$ players represented by
an adjacency matrix $\mathbf{G}(N)$; where $g_{ij}=1$ indicates a link
between players $i$ and $j$, and $g_{ij}=0$ otherwise. Since the adjacency
matrix $\mathbf{G}(N)$ is symmetric and non-negative it follows that its
eigenvalues are real and the maximum eigenvalue $\lambda _{\max }(N)$ is
positive and dominates in magnitude all other eigenvalues.

A complete  bipartite network is a network $\mathbf{g}=(K,M,E)$ of $%
N=\{1,2,...,n\}$ nodes such that the set $N$ can be divided into two
disjoint sets $K,M\subseteq N,$ satisfying that $N=K\cup M$ and $g_{ij}=0$
if $i$ and $j$ belong to the same set ($K$ or $M$) and $g_{ij}=1$ otherwise. 
$E$ is the set of edges. For any coalition of workers $S\subseteq N$; let $%
\mathbf{g}(S),$ denote the subnetwork induced by $S$; with adjacency matrix $%
\mathbf{G}(S)$, and $\lambda _{\max }(S)$ is its maximum eigenvalue. For any
coalition $S\subseteq N$ we can rewrite it as $S=K(S)\cup M(S)$ with $%
K(S):=S\cap K\subseteq K$ and $M(S):=S\cap M\subseteq M$ disjoint sets, and $E(S)$ the set of
edges of coalition $S$. We denote $\left\vert K(S)\right\vert \ $by $k_{S}$
and $\left\vert M(S)\right\vert $ by $m_{S}$ for simplicity.

\section{Finite attenuation network games}

\bigskip In order to facilitate the reader's understanding, we consider a
real context of application of our study. We focus on a firm  where $%
N=\{1,2,...,n\}=K\cup M$ is the total set of workers and $K,M$ two different
groups of fully connected workers. Formally, we consider a 
complete   bipartite network $\mathbf{g}=(K,M,E)$. For any subset/team of workers $%
S\subseteq N$, we know that the induced network is a complete bipartite network  $%
\mathbf{g}(S)=(K(S),M(S),E(S)).$ Consider $t\geq 0$ as a natural number and $%
\delta \geq 0$ as a real number. We define the matrix%
\begin{equation*}
M^{t}(\mathbf{g}(S),\delta )=\sum_{u=0}^{t}\delta ^{u}\mathbf{G}^{u}(S)
\end{equation*}

Note that each entry $m_{ij}^{t}(\mathbf{g}(S),\delta )=\sum_{u=0}^{t}\delta
^{u}\mathbf{g}_{ij}^{u}(S)$ counts the number of walks of at most distance $%
t $ in $\mathbf{g}(S)$ that start in $i$ and end at $j$ weighted by $\delta
^{u}$. In interpretation, the non-negative parameter $\delta $ is an
attenuation factor that scales down the relative weight of longer walks.
Hence, $M^{0}(\mathbf{g}(S),\delta )=\mathbf{I}_{\left\vert S\right\vert
x\left\vert S\right\vert }$ because of $\mathbf{G}^{0}(S)$ is the identity
matrix.

Given a  team $S,$ each worker $i\in S$ has an intrinsic
productivity of $1$ and an actual productivity $p_{i}^{S}(\delta ,t)$ that
benefits from the productivity of the other workers in the team at a
distance of at most $t$ (finite attenuation) in $\mathbf{g}(S),$ at a rate of 
$\delta $. That is:%

\begin{equation*}
p_{i}^{S}(\delta ,t):=\sum_{j\in S}m_{ij}^{t}(\mathbf{g}(S),\delta )
\end{equation*}


Note that  $p_{i}^{S}(\delta ,0)=1$ and  $p_{i}^{S}(\delta ,t)$ for $t>1$ is a measure of the productivity of the worker $i$ in team $S$ that taking into account a peer effects of workers in the team.
\medskip

Now, given a network $\mathbf{g}=(K,M,E)$ we define the corresponding finite
distance attenuation network game (henceforth FAN game) as $(N,v_{\delta
}^{t})$ with $N=K\cup M$ and $t,\delta \geq 0,$ where $v_{\delta
}^{t}(S):=\sum_{i\in S}p_{i}^{S}(\delta ,t)$ for all coalition $S\subseteq
N. $ Note that the characteristic function $v_{\delta }^{t}$ represents the
aggregate productivity of the worker team $S$ up to distance at most $t$
weighted by $\delta $. \bigskip

The following proposition shows that we can explicitly compute the
characteristic function of the FAN games.

\begin{proposition}
Let $\mathbf{g}=(K,M,E)$ be a complete  bipartite network and $(N,v_{\delta
}^{t})$ the corresponding FAN game. For each coalition $S\subseteq N$ it
holds: 
\begin{equation*}
v_{\delta }^{t}(S)=\left\{ 
\begin{array}{ccc}
\left\vert S\right\vert & \text{if} & t=0, \\ 
&  &  \\ 
\left\vert S\right\vert +\left( \left\vert S\right\vert \delta +2\right) 
\overset{\frac{t}{2}}{\underset{u=1}{\sum }}k_{S}^{u}m_{S}^{u}\delta ^{2u-1},
& \text{if} & t\text{ is even}. \\ 
&  &  \\ 
\left\vert S\right\vert +\left( \left\vert S\right\vert \delta +2\right) 
\overset{\frac{t-1}{2}}{\underset{u=1}{\sum }}\left(
k_{S}^{u}m_{S}^{u}\delta ^{2u-1}\right) +2k_{S}^{\frac{t+1}{2}}m_{S}^{\frac{%
t+1}{2}}\delta ^{t}, & \text{if} & t\text{ is odd}.%
\end{array}%
\right.
\end{equation*}%
\newline
\end{proposition}

The reader may notice that $v_{\delta }^{t}(S)>0$ for all $%
S\subseteq N$ and $t,\delta \geq 0$. The increase in productivity with
respect to the increase in distance can be seen more clearly if we relate
FAN games at different distances:%
\begin{eqnarray*}
v_{\delta }^{0}(S) &=&\left\vert S\right\vert \\
v_{\delta }^{1}(S) &=&v_{\delta }^{0}(S)+2k_{S}m_{S}\delta \\
v_{\delta }^{2}(S) &=&v_{\delta }^{1}(S)+\left(
k_{S}^{2}m_{S}+k_{S}m_{S}^{2}\right) \delta ^{2} \\
v_{\delta }^{3}(S) &=&v_{\delta }^{2}(S)+2k_{S}^{2}m_{S}^{2}\delta ^{3} \\
&&\vdots \\
v_{\delta }^{t}(S) &=&\left\{ 
\begin{array}{ccc}
v_{\delta }^{t-1}(S)+\left\vert S\right\vert k_{S}^{\frac{t}{2}}m_{S}^{\frac{%
t}{2}}\delta ^{t}, & \text{if} & t\text{ is even}, \\ 
&  &  \\ 
v_{\delta }^{t-1}(S)+2k_{S}^{\frac{t+1}{2}}m_{S}^{\frac{t+1}{2}}\delta ^{t},
& \text{if} & t\text{ is odd}.%
\end{array}%
\right.
\end{eqnarray*}%

This can be interpreted as follows: when we go from distance $0$
to $1$, each worker (of $K(S)$ or $M(S)$) receives part of the productivity
of the workers of the opposite group, hence the aggregate productivity
increase of the team is $2k_{S}m_{S}\delta =\sqrt{k_{S}m_{S}}\cdot 2\cdot
\left( \sqrt{k_{S}m_{S}}\delta \right) ${. When the distance increases to $2$%
, in addition to the above productivity $(v_{\delta }^{1}(S))$, each worker
also has access to the productivity of his own group for each worker of the
opposite group, and so the increase of the team is now $\left(
k_{S}^{2}m_{S}+k_{S}m_{S}^{2}\right) \delta ^{2}=$}$\frac{k_{S}+m_{S}}{2}%
\cdot 2\left( \sqrt{k_{S}m_{S}}\delta \right) ^{2}${. However if we increase
the distance to $3$ each worker receives, in addition to the above
productivity $(v_{\delta }^{2}(S))$, the productivity of the other group ($%
K(S)$ or $M(S)$) for each path of distance $2$ that may exist, and now the
increase of the team is $2k_{S}^{2}m_{S}^{2}\delta ^{3}=$ $\sqrt{k_{S}m_{S}}%
\cdot 2\cdot $}$\left( \sqrt{k_{S}m_{S}}\delta \right) ^{3}${and so on.}

Our next objective is to analyse the properties of FAN games. It is easy to
see that when the team of workers increases, we add more productivity to the
team, hence FAN games are monotonic. The natural question that arises is whether the snowball effect in productivity whereby     the returns  of  joining  a coalition of workers   increases  as  the  coalition  grows large occurs in our game  (i.e., FAN games are convex).  The following theorem provides affirmative answer. 

\begin{theorem}
Every FAN game is convex.
\end{theorem}

\bigskip The fact that any FAN game is convex has two important
consequences. FAN games are totally balanced and the Shapley value always
belongs to the core of these games. Next, we illustrate how to calculate different FAN games by changing the distance range $t$, through the analysis of a logistic network with several distribution centers.

\begin{example}
\label{example1} We consider the analysis of a logistic network involving three distribution centers: 1, 2, and 3. Distribution centers 2 and 3 do not have a direct relationship in terms of collaboration or resource exchange in this specific logistic network. Each distribution center can operate independently, and its productivity can be influenced by internal factors such as operational efficiency and service quality. However, distribution center 1 is connected to both distribution center 2 and 3. This indicates that its productivity can be influenced by the collaboration and advancements of both distribution centers. There can be information exchange, service provision, or resource sharing between  distribution center 1 and distribution centers 2 and 3, which benefits the overall productivity.

Additionally, we consider the flow of innovations among the distribution centers measured as a distance. This distance reflects the number of steps it takes for innovations to reach a particular distribution center after being evaluated and filtered by others. If the distance is one, each distribution center has direct access to the innovations of the other centers. For example, distribution center 1 can access the results of 2 and 3. If the distance is two, in addition to the aforementioned access, distribution center 1 will also be able to access its own innovations after they have been evaluated by distribution centers 2 and 3.

In this situation, we assume an attenuation factor of $\frac{1}{2} $  meaning that the productivity of each distribution center is halved with each iteration. This factor represents the diminishing impact of previously shared innovations as they propagate through the network.

Formally, we define a complete  bipartite network with $K=\{1\},M$ $%
=\{2,3\}$ and $\delta =\frac{1}{2}.$ The following table shows the
corresponding FAN game with $\delta =\frac{1}{2}$ and $t\in \{0,1,2,3,10\}$
as shown in Table \ref{table}.%
\vspace*{-0.5cm}
\begin{table}[h]
\begin{equation*}
\begin{array}[t]{|c||c|c|c|c|c|c|}
S & v_{\delta }^{t}(S)\text{ } & v_{\delta }^{0}(S) & v_{\delta }^{1}(S) & 
v_{\delta }^{2}(S) & v_{\delta }^{3}(S) & v_{\delta }^{10}(S) \\ \hline
\{i\} & 1 & 1 & 1 & 1 & 1 & 1 \\ \hline
\{2,3\} & 2 & 2 & 2 & 2 & 2 & 2 \\ \hline
\{1,i\} & \left\{ 
\begin{array}{ccc}
2, & \text{if} & t=0, \\ 
2+6\overset{\frac{t}{2}}{\underset{u=1}{\sum }}\left( \frac{1}{4}\right)
^{u}, & \text{if} & t\text{ is even} \\ 
2+\left( \frac{1}{2}\right) ^{t-1}+6\overset{\frac{t-1}{2}}{\underset{u=1}{%
\sum }}\left( \frac{1}{4}\right) ^{u}, & \text{if} & t\text{ is odd},%
\end{array}%
\right. & 2 & 3 & 3.5 & 3.75 & 3.998 \\ \hline
N & \left\{ 
\begin{array}{ccc}
3, & \text{if} & t=0, \\ 
3+7\overset{\frac{t}{2}}{\underset{u=1}{\sum }}\left( \frac{1}{2}\right)
^{u}, & \text{if} & t\text{ is even}, \\ 
3+\left( \frac{1}{2}\right) ^{\frac{t-3}{2}}+7\overset{\frac{t-1}{2}}{%
\underset{u=1}{\sum }}\left( \frac{1}{2}\right) ^{u}, & \text{if} & t\text{
is odd},%
\end{array}%
\right. & 3 & 5 & 6.5 & 7.5 & 9.78125 \\ \hline
\end{array}
\end{equation*}
\caption{FAN games for $t=0,1,2,3,10$ for Example \ref{example1} \label{table}}
\end{table}

Table \ref{table 1} shows the productivity of each center $i$ in the overall network
for the above flows of innovations (distances).
\begin{table}[h]
\begin{equation*}
\begin{tabular}{|c||c|c|c|c|c|}
$Worker$ & $p_{i}^{N}(\frac{1}{2},0)$ & $p_{i}^{N}(\frac{1}{2},1)$ & $%
p_{i}^{N}(\frac{1}{2},2)$ & $p_{i}^{N}(\frac{1}{2},3)$ & $p_{i}^{N}(\frac{1}{%
2},10)$ \\ \hline
$1$ & $1$ & $2$ & $2.5$ & $3$ & $3.90625$ \\ 
$2$ & $1$ & $1.5$ & $2$ & $2.25$ & $2.9375$ \\ 
$3$ & $1$ & $1.5$ & $2$ & $2.25$ & $2.9375$ \\ \hline
\end{tabular}%
\end{equation*}
\caption{Productivity in $N$ for $t=0,1,2,3,10$ for Example \ref{example1} \label{table 1}}
\end{table}

We may notice that the larger $t$ the higher  individual and aggregate
productivities. Moreover, productivities seem to converge to a certain value
as the flow of innovation $t$ increases, i.e, $p^{N}(\frac{1}{2},t)\approx (4,3,3)$
for $t$ enough large. In conclusion, we can say that distribution center 1 has a higher final productivity than the others.
\end{example}

A question that may arise naturally is  whether FAN games converge to a particular game when $t$
increases. In the following section we determine necessary and sufficient
conditions on attenuation factor $\delta $ for FAN games to converge (when $%
t$ goes to infinity).

\section{Converging FAN games to Attenuation games}

\bigskip In this section we  investigate  what happens when each worker in a team
benefits from the productivity of the others at any distance,
that is, what happens to FAN games when the distance goes to infinity. We
are interested  in study under what conditions FAN games converge to a well-defined TU-game.

Consider a complete  bipartite  network $\mathbf{g}=(K,M,E)$ and $\Lambda
(g,\delta ):=\left\{ (N,v_{\delta }^{t})/t\in 
\mathbb{N}
\right\} $ the family of all possible FAN games with an attenuation factor $%
\delta \geq 0 $. It is easy to check that $\lambda _{\max}(S)=\sqrt{%
k_{S}m_{S}},$ for all $S \subseteq N.$

The first theorem provides a necessary and sufficient condition for the
family of FAN games to converge. Before showing it we need the following
technical lemma.

\begin{lemma}
\label{lemma 1}Let $\mathbf{g}$ be a complete  bipartite network and $\Lambda
(g,\delta )$ the corresponding family of FAN games with $\delta $. Then, $%
\left\{ v_{\delta }^{t}(S)\right\} _{t\in 
\mathbb{N}
}$ converges to a real value $v_{\delta }(S),\ $for each coalition $%
S\subseteq N,$ if and only if $\delta \in \left[ 0,\frac{1}{\lambda _{\max
}(S)}\right[ .$
\end{lemma}

Note that this technical condition sets a different condition for the convergence of the productivity of each team based on the same attenuation factor. The following result provides a unique condition in terms of the network's overall productivity.
\begin{theorem}
Let $\mathbf{g}$ be a complete  bipartite network and $\Lambda (g,\delta )$
the corresponding family of FAN games with $\delta $. Then, $\left\{
v_{\delta }^{t}\right\} _{t\in 
\mathbb{N}
}$ converges to a finite TU game $v_{\delta }$ if and only if $\delta \in %
\left[ 0,\frac{1}{\lambda _{\max }(N)}\right[ .$
\end{theorem}

Given $g$ a complete  bipartite network and the associated family of FAN
games $\Lambda (g,\delta )$ with $\delta \in \left[ 0,\frac{1}{\lambda
_{\max }(N)}\right[ $, we can define an attenuation network game $%
(N,v_{\delta })$ as the limit of $\left\{ v_{\delta }^{t}\right\} _{t\in 
\mathbb{N}
}.$ Notice this game is well defined because of the above theorem.
Henceforth, we will refer to $(N,v_{\delta })$ as a AN game. Moreover, by
lemma \ref{lemma 1}, we have an explicit formula for AN games, that is, for
any $S\subseteq N,$ 
\begin{equation*}
v_{\delta }(S)=\frac{k_{S}+m_{S}+2k_{S}m_{S}\delta }{1-k_{S}m_{S}\delta ^{2}}%
.
\end{equation*}%
The following example illustrates AN games and the distribution of the
individual productivity in the grand  coalition.

\begin{example} \label{example2}
Consider the example \ref{example1} with $K=\{1\},M$ $=\{2,3\}$ and $\delta =%
\frac{1}{2}.$ Notice that $\lambda _{\max }(S)=\sqrt{2}$ and $\delta =1/2\in %
\left[ 0,\frac{1}{\sqrt{2}}\right[ .$

Table \ref{table 2} shows that the limit of the family of FAN games is a TU
game with a finite values

\begin{table}[!h]
\begin{equation*}
\begin{tabular}{|c||c|c|c|c|}
$S$ & $\{i\}$ & $\{2,3\}$ & $\{1,i\}$ & $N$ \\ \hline
$v_{\delta }^{0}(S)$ & $1$ & $2$ & $2$ & $3$ \\ \hline
$v_{\delta }^{1}(S)$ & $1$ & $2$ & $3$ & $5$ \\ \hline
$v_{\delta }^{2}(S)$ & $1$ & $2$ & $3.5$ & $6.5$ \\ \hline
$v_{\delta }^{3}(S)$ & $1$ & $2$ & $3.75$ & $7.5$ \\ \hline
$v_{\delta }^{10}(S)$ & $1$ & $2$ & $3.998$ & $9.78125$ \\ \hline
$\vdots $ & $\vdots $ & $\vdots $ & $\vdots $ & $\vdots $ \\ \hline
$v_{\delta }(S)$ & $1$ & $2$ & $4$ & $10$ \\ \hline
\end{tabular}%
\end{equation*}
\caption{Convergence of the FAN-games for Example \ref{example2} \label{table 2}}
\end{table}

Moreover, the limit of the individual productivity for the grand  coalition, $%
\underset{t\rightarrow \infty }{\lim }p^{N}(\frac{1}{2},t)=(4,3,3):=p^{N}(%
\frac{1}{2})$ is a stable (in the sense of the core) distribution of the
total productivity $(v_{\delta }(N)=10).$
\end{example}

Next proposition shows that $p^{N}(\delta ):=\underset{t\rightarrow \infty }{%
\lim }p^{N}(\delta ,t),$ is always a core allocation for $(N,v_{\delta }).$
Hence, AN games are totally balanced, because of every subgame of an AN game
is also an AN game.

\begin{proposition}
Let $\mathbf{g}$ be a complete  bipartite network and $(N,v_{\delta })$ be
the corresponding AN game. Then, $p^{N}(\delta )\in Core(N,v_{\delta }).$
\end{proposition}

\medskip Next theorem proves that AN games are convex. Before introducing
it, let's demostrate the following technical lemma, which shows the marginal
productivity of a worker to a team.

\begin{lemma}
Let $\mathbf{g}$ be a complete  bipartite network and $(N,v_{\delta })$ be
the corresponding AN game. Then, for any $i \in S \subseteq N,$\newline
\begin{equation*}
v_{\delta }(S)-v_{\delta }(S\backslash \{i\})=\left\{ 
\begin{array}{ccc}
\frac{\left( 1+m_{S}\delta \right) ^{2}}{(1-k_{S}m_{S}\delta
^{2})(1-k_{S}m_{S}\delta ^{2}+m_{S}\delta ^{2})}, & \text{if} & i\in K(S),
\\ 
&  &  \\ 
\frac{\left( 1+k_{S}\delta \right) ^{2}}{(1-k_{S}m_{S}\delta
^{2})(1-k_{S}m_{S}\delta ^{2}+k_{S}\delta ^{2})}, & \text{if} & i\in M(S).%
\end{array}%
\right.
\end{equation*}
\end{lemma}

The following theorem shows that the marginal productivity of a worker to a
team is greater the larger the team is.\footnote{It is worth noting that     convexity  of AN games  can be also shown to  follow, via  a limit argument,  from the convexity of the  FAN games}
\begin{theorem}
Every AN game is convex.
\end{theorem}

\bigskip As mencioned above, the Shapley value, $\phi (v_{\delta }),$ always
belongs to the core of the AN game $(N,v_{\delta })$. Next theorem provides
a explicit formula for the Shapley value of AN games.

\begin{theorem}
Let $g$ a complete bipartite network and $(N,v_{\delta })$ the corresponding
AN game. Then, for all $i\in K$\newline
\newline
\begin{equation*}
\phi _{i}(v_{\delta })=\sum\limits_{k=1}^{\left\vert K\right\vert
}\sum\limits_{m=0}^{\left\vert M\right\vert }\Pi ^{K}_{M}(k,m)\cdot \frac{%
\left( 1+m\delta \right) ^{2}}{(1-km\delta ^{2})(1-km\delta ^{2}+m\delta
^{2})}
\end{equation*}%
\newline
and for all $i\in M$\newline
\begin{equation*}
\phi _{i}(v_{\delta })=\sum\limits_{k=0}^{\left\vert K\right\vert
}\sum\limits_{m=1}^{\left\vert M\right\vert }\Pi ^{M}_{K}(m,k)\cdot \frac{%
\left( 1+k\delta \right) ^{2}}{(1-km\delta ^{2})(1-km\delta ^{2}+k\delta
^{2})}
\end{equation*}%
\newline
\newline
where $\Pi ^{X}_{Y}(i,j)=\binom{\left\vert Y\right\vert }{j}\cdot \binom{%
\left\vert X\right\vert -1}{i-1}\cdot \frac{\left( i+j-1\right) !(\left\vert X\right\vert +\left\vert Y\right\vert -i-j)!}%
{\left(\left\vert X\right\vert+\left\vert Y\right\vert\right)!%
}$
\newline
\newline
\end{theorem}

The reader may notice that once we obtain the Shapley value for a worker $%
i\in K$, $\phi _{i}(v_{\delta }),$ it is easy to calculate it for workers $%
j\in M.$ Indeed, $\phi _{j}(v_{\delta })=\frac{v_{\delta }(N)-\left\vert
K\right\vert \cdot \phi _{i}(v_{\delta })}{\left\vert M\right\vert }$ for
all $j\in M$ and $i\in K.$

 Recall that, as we already discussed earlier, the overall productivity can be considered a public good. The Shapley value acts then as an individual measure for productivity. Additionally, the Shapley value can be interpreted as an individual's contribution to the public good, demonstrating a voluntary willingness to contribute to the sustainability of that shared productivity. Next example illustrate the Shapley value for AN games.

\begin{example}\label{example3}
Consider again the example \ref{example1} with $K=\{1\},M$ $=\{2,3\}$ and $%
\delta =\frac{1}{2}.$ Table \ref{table 3} compares the Shapley value with
individual productivity for the grand coalition.

\begin{table}[!h]
\begin{equation*}
\begin{tabular}{|c||c|c|}
Worker & $p^{N}(\frac{1}{2})$ & $\phi (v_{\frac{1}{2}})$ \\ \hline
$1$ & $4$ & $4$ \\ 
$2$ & $3$ & $3$ \\ 
$3$ & $3$ & $3$ \\ \hline
\end{tabular}%
\end{equation*}
\caption{Productivity in $N$ vs Shapley value for Example \ref{example3} \label{table 3}}
\end{table}

In this example, both productivity distributions coincides but this is not
the case in general. After an extended interaction among the centers,  center 1 contributes to the network with a productivity level of 4, while the rest of the centers contribute with a level of 3 each.
\end{example}

The following example shows that Shapley value can be close to the
individual productivity for the grand coalition.

\begin{example}\label{example4}
Consider the logistic network given in the example \ref{example1} expanded with a new distribution center 4, that is, $K=\{1\},M$ $=\{2,3,4\}$, but
now $\delta =\frac{1}{3}.$ Notice that $\lambda _{\max}(N)=\sqrt{3}$ and $%
\delta=1/3 \in \left[ 0,\frac{1}{\sqrt{3}}\right[.$

Table \ref{table4} compares the Shapley value with individual productivity
for the grand coalition. 

\begin{table}[!h]
\begin{equation*}
\begin{tabular}{|c||c|c|}
Worker & $p^{N}(\frac{1}{3})$ & $\phi (v_{\frac{1}{3}})$ \\ \hline
$1$ & $3$ & $3.14$ \\ 
$2$ & $2$ & $1.95$ \\ 
$3$ & $2$ & $1.95$ \\ 
$4$ & $2$ & $1.95$ \\ \hline
\end{tabular}%
\end{equation*}
\caption{Productivity in $N$ vs Shapley value for Example \ref{example4} \label{table4}}
\end{table}

The reader may notice that center 1 has an individual productivity level of 3, but contributes to the network with a productivity level of 3.14. On the other hand, the rest of the centers have an individual productivity level of 2, while their contribution to the network is lower (1.95).
\end{example}

Notice that while $p_{i}^{N}(\delta )$ represents the individual
productivity of worker $i$ in the network, $\phi _{i}(v_{\delta })$ is
interpreted as the average marginal productivity of such a worker $i$ in all
the teams. However, despite having an explicit formula for the
Shapley value, it is still difficult to calculate when the number of workers
grows. Moreover, we observe that in AN games with few workers the two are
close, matching in some cases. Next, we focus on finding an alternative
productivity distribution for AN games which takes into account the increase
in productivity in the distance of the grand coalition, as well as the the
degree of connectivity of each worker.

\section{Productivity distribution that recognizes workers' connectivity}

We first go back to FAN games and study in detail what happens when the
distance increases. It is important to measure how much productivity each
team generates as the distance $t$ increases. This information will allow us
to define an alternative productivity distribution for AN games which, unlike Shapley value, takes into consideration the degree of connectivity of workers.

We start by defining the difference game in $t$, as the difference between
FAN games in $t$ and $t-1$. Formally, $(N,d_{\delta }^{t})$ such that for
all coalition $S\subseteq N$, $d_{\delta }^{t}(S):=v_{\delta
}^{t}(S)-v_{\delta }^{t-1}(S)$.

Next proposition shows and explicit formula for the difference games in $t$. 

\begin{proposition}
Let $\mathbf{g}$ be a complete  bipartite network and $\Lambda (\delta
):=\left\{ (N,v_{\delta }^{t})/t\in 
\mathbb{N}
\right\} $ the family of FAN games. Then, difference games $(N,d_{\delta
}^{t})$ with $t\geq 0,$ are given by,%
\begin{equation*}
d_{t}^{\delta }(S)=\frac{1}{2}\left[ \left( \sqrt{k_{S}}+\sqrt{m_{S}}\right)
^{2}\left( \lambda _{\max }(S)\right) ^{t}+\left( \sqrt{k_{S}}-\sqrt{m_{S}}%
\right) ^{2}\left( -\lambda _{\max }(S)\right) ^{t}\right] \ \delta ^{t}
\end{equation*}%
for all $S\subseteq N.$
\end{proposition}

Notice that the difference game for a distance $t$, $(N,d_{\delta }^{t}),$
measures the increase in productivity at FAN games per unit of distance.
That is, the increase in productivity from $(N,v_{\delta }^{t-1})$ to $%
(N,v_{\delta }^{t})$. Moreover, we can rewrite $d_{\delta }^{t} $ as follows:%
\begin{equation*}
d_{t}^{\delta }(S)=\left[ \left( \frac{k_{S}+m_{S}}{2}+\sqrt{k_{S}m_{S}}%
\right) \left( \lambda _{\max }(S)\right) ^{t}+\left( \frac{k_{S}+m_{S}}{2}-%
\sqrt{k_{S}m_{S}}\right) \left( -\lambda _{\max }(S)\right) ^{t}\right] \
\delta ^{t}
\end{equation*}

Thus, we can distinguish that in even periods the increase in productivity
is influenced by the arithmetic mean, $\frac{k_{S}+m_{S}}{2},$ while in odd
periods it is influenced by the geometric mean, $\sqrt{k_{S}m_{S}}$. Given
that, $\frac{k_{S}+m_{S}}{2}\geq \sqrt{k_{S}m_{S}}$ for all team $S\subseteq
N$. We can deduce that productivity increases more when we
extend the possibility for workers to obtain productivity from odd to even
distance than vice versa. This effect is due to the complete bipartite
structure of the network as mentioned in the previous section.

Based on this definition we can also rewrite the game $v_{\delta }^{t}$ in
the following way

\begin{equation*}
v_{\delta }^{t}(S)=\left\vert S\right\vert +\frac{1}{2}\sum_{u=1}^{t}\left( %
\left[ \left( \sqrt{k_{S}}+\sqrt{m_{S}}\right) ^{2}\left( \lambda _{\max
}(S)\right) ^{u}+\left( \sqrt{k_{S}}-\sqrt{m_{S}}\right) ^{2}\left( -\lambda
_{\max }(S)\right) ^{u}\right] \ \delta ^{u}\right)
\end{equation*}

We use now the structure of difference games to define a productivity
distribution for AN games. Given $d_{\delta }^{t}(N)$ for any $t\geq 1$,
we define a productivity distribution $x^{t}(\delta )=(x_{i}^{t}(\delta
))_{i\in N}$ such that 

\begin{equation*}
x_{i}^{t}(\delta )=\left\{ 
\begin{array}{ccc}
\frac{d_{\delta }^{t}(N)}{\left\vert N\right\vert }\cdot \frac{\left\vert M\right\vert }{\left\vert%
K\right\vert }, & \text{if} & i\in K \\ 
&  &  \\ 
\frac{d_{\delta }^{t}(N)}{\left\vert N\right\vert }\cdot \frac{\left\vert K\right\vert}{\left\vert%
M\right\vert}, & \text{if} & i\in M%
\end{array}%
\right.
\end{equation*}

We may notice that we first divide the total
productivity among all workers equally, then we weight it by the ratio
between the number of $K$ and $M$ nodes. So workers in set $K$ receive more
if the number of links leaving each worker ($\left\vert M\right\vert $) is
greater than those of the workers in $M$ ($\left\vert K\right\vert $) and
vice versa.

Next proposition shows that $x^{t}(\delta)$ is stable in the sense of the
core.

\begin{proposition}
Let $\mathbf{g}$ be a complete  bipartite network and $(N,d_{\delta }^{t})$
the difference game in $t$. Then, $x^{t}(\delta ) \in Core(N,d_{\delta
}^{t}).$
\end{proposition}

The following example illustrates the difference games for distances $%
t\in\left\{1,2,3,4,5\right\}$.

\begin{example} \label{example5}
Consider again the example \ref{example1} with $K=\{1\},M$ $=\{2,3\}$ and $%
\delta =\frac{1}{2}.$ Table \ref{table 6} shows the difference games for $t\in
\{1,2,3,4,5\}$
\vspace*{-0.5cm}
\begin{table}[!h]
\begin{equation*}
\begin{array}[t]{|c||c|c|c|c|}
S & \{i\}\text{ } & \{2,3\} & \{1,i\} & N \\ \hline
d_{\delta }^{t}(S)\text{ } & 1 & 2 & \left( \frac{1}{2}\right) ^{t-1} & 
\left[ \left( \frac{3}{2}+\sqrt{2}\right) \left( \sqrt{2}\right) ^{t}+\left( 
\frac{3}{2}-\sqrt{2}\right) \left( -\sqrt{2}\right) ^{t}\right] \left( \frac{%
1}{2}\right) ^{t} \\ \hline
d_{\delta }^{1}(S) & 0 & 0 & 1 & 2 \\ \hline
d_{\delta }^{2}(S) & 0 & 0 & 0.5 & 1.5 \\ \hline
d_{\delta }^{3}(S) & 0 & 0 & 0.25 & 1 \\ \hline
d_{\delta }^{4}(S) & 0 & 0 & 0.125 & 0.75 \\ \hline
d_{\delta }^{5}(S) & 0 & 0 & 0.0625 & 0.5 \\ \hline
\end{array}%
\end{equation*}
\caption{Difference games for Example \ref{example5} \label{table 6}}
\end{table}


Table \ref{table 7} shows the calculation of the productivity distribution $%
x_{i}^{t}(\delta )$ for distances $t\in\left\{1,2,3,4,5\right\}$


\begin{table}[!h]
\begin{equation*}
\begin{tabular}{|c||c|c|c|c|c|}
Worker & $x_{i}^{1}(\delta )$ & $x_{i}^{2}(\delta )$ & $x_{i}^{3}(\delta )$
& $x_{i}^{4}(\delta )$ & $x_{i}^{5}(\delta )$ \\ \hline
1 & $4/3$ & $1$ & $2/3$ & $1/2$ & $1/3$ \\ \hline
2 & $1/3$ & $1/4$ & $1/6$ & $1/8$ & $1/12$ \\ \hline
3 & $1/3$ & $1/4$ & $1/6$ & $1/8$ & $1/12$ \\ \hline
\end{tabular}%
\end{equation*}
\caption{Distribution $x^{t}(\delta )$ for $t=1,...,5$ for Example \ref{example5} \label{table 7}}
\end{table}
\end{example}

We are now ready to build a productivity distribution for AN games based on
the difference distribution $x^{t}(\delta)$. Consider $\mathbf{g}$ a
bipartite complete network and $(N,v_{\delta })$ its corresponding AN game.
We define the link ratio productivity distribution (henceforth LRP
distribution) as the equal distribution of the increase in productivity ($%
\frac{d_{\delta }^{t}(N)}{N})$ with respect to the link ratio ($\frac{%
\left\vert M\right\vert }{\left\vert K\right\vert }$ or $\frac{\left\vert
K\right\vert }{\left\vert M\right\vert }$ depending of the worker
considered). Formally, it is constructed by adding to $1$ (the individual
productivity) the sum of the difference distributions $x_{i}^{t}(\delta )$
of each distance $t\geq1$, that is

\begin{equation*}
\omega (\delta ):=\mathbf{1}_{N}+\underset{t\rightarrow \infty }{\lim }%
\left( \overset{t}{\underset{u=1}{\sum }}x^{u}(\delta )\right).
\end{equation*}

Notice that, when $\delta =0, d_{\delta }^{t}(S)=0,$ for all $S \subseteq N,$
then $x^{t}(\delta)=0_{N}$ and so, $\omega (\delta )=\mathbf{1}_{N}.$
\medskip

Next proposition provides an explicit formula for LRP distribution when $%
\delta >0$.

\begin{proposition}
Let $\mathbf{g}$ be a complete  bipartite network an $(N,d_{\delta }^{t})$
the corresponding difference games for $t\geq1$ and $\delta>0$. Then, the
LRP distribution $\omega (\delta )$ is given by:%
\begin{equation*}
\omega _{i}(\delta )=\left\{ 
\begin{array}{ccc}
1+\left( \frac{\left\vert M\right\vert }{\left\vert K\right\vert }\delta +%
\frac{2\left\vert M\right\vert }{\left\vert N\right\vert \left\vert
K\right\vert }\right) \frac{\left\vert K\right\vert \left\vert M\right\vert
\delta }{1-\left\vert K\right\vert \left\vert M\right\vert \delta ^{2}}, & 
\text{if} & i\in K, \\ 
&  &  \\ 
1+\left( \frac{\left\vert K\right\vert }{\left\vert M\right\vert }\delta +%
\frac{2\left\vert K\right\vert }{\left\vert N\right\vert \left\vert
M\right\vert }\right) \frac{\left\vert K\right\vert \left\vert M\right\vert
\delta }{1-\left\vert K\right\vert \left\vert M\right\vert \delta ^{2}}, & 
\text{if} & i\in M.%
\end{array}%
\right.
\end{equation*}
\end{proposition}

The following theorem shows that LRP distribution is stable in the sense of
the core.

\begin{theorem}
Let $\mathbf{g}$ be a complete bipartite network and $(N,v_{\delta }) $ the
corresponding AN game. Then, $\omega (\delta ) \in Core(N,v_{\delta })$.
\end{theorem}

To conclude this section, we present a characterization of the LRP
distribution. It is based on three appealing properties for AN games. The
first one, \textit{Efficiency} means that the total benefit is divided among
the workers. The second, \textit{equality in bipartition} ensures that all
workers originating the same number of links have the same productivity
distribution. The last one, \textit{link balanced productivity} property
shows that the productivity of the workers in $K$, discounting their
individual productivity, divided by the average number of links, is exactly
equal to the workers in $M$. This guarantees an equal contribution of each
link to the productivity of the network.

Formally, we consider a network $\mathbf{g}$ and the corresponding AN game $%
(N,v_{\delta }).$ We define the following three properties for a
single-valued solution $\varphi $ on AN games $(N,v_{\delta })$:

\begin{description}
\item[(EF)] \textit{Efficiency}. $\sum_{i\in N}\varphi _{i}(v_{\delta
})=v_{\delta }(N).$

\item[(EB)] \textit{Equality in bipartition}. $\varphi _{i}(v_{\delta
})=\varphi _{j}(v_{\delta })$ for all $i,j\in K$ and $\varphi _{i}(v_{\delta
})=\varphi _{j}(v_{\delta })$ for all $i,j\in M.$

\item[(LBP)] \textit{Link balanced productivity}. $\frac{1}{\left\vert
M\right\vert }\sum_{i\in K}\left( \varphi _{i}(v_{\delta })-1\right) =\frac{1%
}{\left\vert K\right\vert }\sum_{j\in M}\left( \varphi _{j}(v_{\delta
})-1\right) .$
\end{description}

The last theorem in this paper states that there exists a unique
productivity distribution for AN games satisfying the properties EF, EB and
LBP.

\begin{theorem}
\label{charac}Let $\mathbf{g}$ be a complete  bipartite network and $%
(N,v_{\delta })$ the corresponding AN game. Then, the LRP sitribution $%
\omega (\delta )$. is the unique productivity distribution satisfying EF, EB
and LBP.
\end{theorem}

The following examples compare the individual productivity distribution, the
Shapley value and de LRP distribution.

\begin{example}\label{example8}
Consider again the example \ref{example1} with $K=\{1\},M$ $=\{2,3\}$ and $%
\delta =\frac{1}{2}$. Table \ref{table8} compares the LRP distribution with
the Shapley value and the individual productivity distribution in the grand
coalition.

\begin{table}[!h]
\begin{equation*}
\begin{array}{|c||c|c|c|}
Worker & p^{N}(\delta ) & \phi (v_{\delta }) & \omega (\delta ) \\ \hline
1 & 4 & 4 & 17/3 \\ 
2 & 3 & 3 & 13/6 \\ 
3 & 3 & 3 & 13/6 \\ \hline
\end{array}%
\end{equation*}
\caption{Shapley value vs LRP distribution for Example \ref{example8} \label{table8}}
\end{table}
\end{example}

\begin{example}\label{example9}
Consider the AN game $(N,v_{\delta })$ with $K=\{1\}$ and $M=\{2,3,4\},$
as shown in Table \ref{table10}.

\begin{table}[!h]
\begin{equation*}
\begin{array}{|c||c|c|c|}
S & v_{\delta }(S)\text{ } & v_{\frac{1}{2}}(S) & v_{\frac{1}{3}}(S) \\ 
\hline
\{i\} & 1 & 1 & 1 \\ \hline
\begin{array}[t]{c}
\{2,3\} \\ 
\{2,4\} \\ 
\{3,4\}%
\end{array}
& 2 & 2 & 2 \\ \hline
\{1,i\} & \frac{2}{1-\delta } & 4 & 3 \\ \hline
\{2,3,4\} & 3 & 3 & 3 \\ \hline
\begin{array}{c}
\{1,2,3\} \\ 
\{1,2,4\} \\ 
\{1,3,4\}%
\end{array}
& \frac{3+4\delta }{1-2\delta ^{2}} & 10 & \frac{39}{7} \\ \hline
N & \frac{4+6\delta }{1-3\delta ^{2}} & 28 & 9 \\ \hline
\end{array}%
\end{equation*}
\caption{AN games for Example \ref{example9} \label{table10}}
\end{table}

Table \ref{table11} compares LRP distribution with the Shapley value and the
individual productivity distribution in the grand coalition for two different
values of $\delta$.
\begin{table}[!h]
\begin{equation*}
\begin{array}{|c||c|c|c||c|c|c|}
Worker & p^{N}\left( \frac{1}{2}\right) & \phi (v_{\frac{1}{2}}) & \omega (%
\frac{1}{2}) & p^{N}(\frac{1}{3}) & \phi (v_{\frac{1}{3}}) & \omega (\frac{1%
}{3}) \\ \hline
1 & 10 & 9.25 & 19 & 3 & 3.14 & 4.75 \\ 
2 & 6 & 6.25 & 3 & 2 & 1.95 & 1.41 \\ 
3 & 6 & 6.25 & 3 & 2 & 1.95 & 1.41 \\ 
4 & 6 & 6.25 & 3 & 2 & 1.95 & 1.41 \\ \hline
\end{array}%
\end{equation*}
\caption{Productivity, Shapley value and LRP distribution for Example \ref{example9} \label{table11}}
\end{table}
\end{example}

Both examples show how worker $1$ gets higher productivity as all links come
out of him, but LRP distribution allocates a higher productivity than the
Shapley value. In other words, if he leaves the network, the other workers
would be disconnected. The LRP distribution compensates much more for the
role of worker 1 in network connectivity.

The reader may notice that If $\left\vert K\right\vert =\left\vert
M\right\vert \ $by efficiency $\phi (v_{\delta })=p^{N}(\delta )=$ $\omega
(\delta )$. If $\left\vert K\right\vert \neq \left\vert M\right\vert ,$ LRP
distribution     assigns higher productivity to those workers who have a higher
number of links, recognizing their greater contribution to the
interconnectedness of the network.

While the Shapley value is an effective measure for  the weighted marginal productivity contribution of a node to various teams,  the LRP distribution serves a distinct role in evaluating the productivity of the entire network in terms of its connections.  Notably, the LRP distribution holds the advantage of being easier to calculate than the Shapley value  for  a complete bipartite network.  To illustrate this, consider example \ref{example9}, where Node 1 emerges as more pivotal in the LRP due to its central role as the starting point for all network connections. In    scenarios where the objective is  to assess   the marginal contribution of workers to different work teams, the Shapley value proves to be a valuable indicator.

In a network context,  readers could consider employing   other off-the-shelf  centrality measures   to establish a ranking for different workers.  It is important  to note that nodes in sets $K$ or $M$ are indistinguishable in terms of centrality measures. Consequently, regardless of the choice of centrality measures, they would not aid in distinguishing between  nodes in sets $K$ or $M$. Moreover,  the various allocations proposed in this work can also be viewed as centrality measures, as they are derived from distinct characteristics of nodes in each set to determine their values.

Finally, we prove that properties used in Theorem \ref{charac} are logically
independent.

\begin{example}
(LBP fails) Consider $\varphi $ on AN game $(N,v_{\delta })$ defined by $%
\varphi (v_{\delta }):=p^{N}(v_{\delta })$\noindent\ where $\left\vert
K\right\vert =1$, $\left\vert M\right\vert =2$ and $\delta =\frac{1}{2}.$
\noindent $\varphi (v_{\delta })$ satisfies EF, EB, but not LBP since $\frac{%
1}{2}\sum_{i\in k}\left( 4-1\right) =\frac{3}{2}\neq 4=\sum_{i\in M}\left(
3-1\right). $
\end{example}

\begin{example}
(EB fails) Consider $\varphi $ on AN game $(N,v_{\delta })$ given by $%
\varphi (v_{\delta }):=(0,2,0,2)$ where $K=\{1,2\}$, $M=\{3,4\}$ and $\delta
=0.$ $\varphi (v_{\delta })${\ } satisfies EF, LBP but not EB.
\end{example}

\begin{example}
(EF fails) Let $\varphi $ on AN game $(N,v_{\delta })$ defined by $\varphi
(v_{\delta }):=p^{N}(v_{\delta })-\mathbf{1}_{N}$\noindent\ where $%
\left\vert K\right\vert =2$, $\left\vert M\right\vert =2$ and $\delta =\frac{%
1}{2}.$ \noindent $\varphi (v_{\delta })${\ } satisfies LBP, EB but not EF.
\end{example}

\section{Concluding remarks}

Network productivity can be considered a public good  in the context of providing delivering common goods in areas such as the environment, health, and logistics. Its role in accessibility, interconnection, positive externalities, and the need for public-private collaboration supports the notion that network productivity is a crucial component for the effective provision of common goods for the benefit of society as a whole.

In this paper, we  have explored both the theory of cooperative games and networks in the
context of productivity measures in logistics infrastructure,  where two teams
of agents/workers interact. We have focused on the structure of complete
bipartite networks because they possess an interesting structural feature
from the point of view of cooperative game theory, i.e., any subnetwork induced  by a coalition of workers
maintains the same  structure and properties of the original network,
allowing the results obtained to be applicable both to the whole
infrastructure and to small teams of workers. From this synergy between
networks and cooperative games arise finite attenuation network
games (FAN games) and attenuation network games (AN games). We have shown
that FAN games converge to AN games  for attenuation factors below a
certain threshold. Then, we have   considered a coalitionally stable productivity
distribution   of the   overall  productivity of the network.  In addition, we have provided an explicit formula  of the Shapley
value and explored an alternative productivity distribution, LRP
distribution, which is easier to compute than the Shapley value, and lends itself nicely to the underlying network structure of interactions. Finally, we have characterised this distribution on
the basis of three properties suitable for a realistic and
functional network.

This work has implications for both academics and practitioners in this
field.  It is crucial to underscore that we have utilized the distinctive structure of complete bipartite networks to derive explicit formulas for both defining the games and proposing allocations. A promising avenue for future research would be to expand this investigation to more general network structures such as  exploring complete multipartite networks or nested split networks. It could serve as a natural extension of our current work. Other future research could further explore the properties of AN games and
their applications in various contexts. Overall, this paper contributes
to the understanding of cooperative game theory and networks, and provides
insights for the design and management of networks with peer effects and  cooperative
objectives. Finally, we propose more specific future research from the
perspective of game theory, such as: (1) Analyzing the differences between
the Shapley value and the individual productivity distribution in the grand
coalition; (2) Extending the study to complete multipartite networks; (3) Finding alternative productivity sharing methods based on
other structural features or properties  of  the network; (4) Analyzing other models in which
the productivity of each worker depends on different types of local interactions and peer effects.

\subsection*{Acknowledgements}
We are grateful to two anonymous referees for helpful comments. This work is part of the R+D+I project grant PID2022-137211NB-100 that was funded by MCIN/AEI/10.13039/50110001133/ and by "ERDF A a way of making EUROPE/UE". This research was also funded by project PROMETEO/2021/063 from the Comunidad Valenciana.

\subsection*{Declarations}
\textbf{Conflict of interest} Not applicable.

\appendix

\section{Appendix}

\begin{proof}[Proof of Proposition 3.1]
Consider a $(K,M,E)$ complete  bipartite network. Take $S\subseteq N$, and
the corresponding subnetwork $g(S)=(K(S),M(S),E(S))$, with matrix:%
\begin{equation*}
\mathbf{G}(S)=\left( 
\begin{array}{cc}
\text{{\Large 0}}_{k_{S}xk_{S}} & \text{{\Large 1}}_{k_{S}xm_{S}} \\ 
\text{{\Large 1}}_{m_{S}xk_{S}} & \text{{\Large 0}}_{m_{S}xm_{S}}%
\end{array}%
\right) _{\left\vert S\right\vert x\left\vert S\right\vert }.
\end{equation*}

$\mathbf{G}^{u}(S)$ can be easily calculated. Indeed, if $u$ is an even
number, then $u=2d$ with $d$ a natural number. Then, 
\begin{equation*}
\mathbf{G}^{2d}(S)=\left( 
\begin{array}{cc}
\left( k_{S}^{d-1}\cdot m_{S}^{d}\right) _{k_{S}xk_{S}} & \text{{\Large 0}}%
_{k_{S}xm_{S}} \\ 
\text{{\Large 0}}_{m_{S}xk_{S}} & \left( k_{S}^{d}\cdot m_{S}^{d-1}\right)
_{m_{S}xm_{S}}%
\end{array}%
\right) _{\left\vert S\right\vert x\left\vert S\right\vert }
\end{equation*}

If, on the other hand, it is odd $u=2d+1$ 
\begin{equation*}
\mathbf{G}^{2d+1}(S)=\left( 
\begin{array}{cc}
\text{{\Large 0}}_{k_{S}xk_{S}} & \left( k_{S}^{d}\cdot m_{S}^{d}\right)
_{k_{S}xm_{S}} \\ 
\left( k_{S}^{d}\cdot m_{S}^{d}\right) _{m_{S}xk_{S}} & \text{{\Large 0}}%
_{m_{S}xm_{S}}%
\end{array}%
\right) _{\left\vert S\right\vert x\left\vert S\right\vert }
\end{equation*}

The expression of $m_{ij}^{t}(\mathbf{g}(S),\delta )=\sum_{u=0}^{t}\delta
^{u}\mathbf{g}_{ij}^{u}(S)$ varies depending on which set of the bipartite
graph the players are located in, we distinguish the following cases:

\begin{itemize}
\item If $i\in K,$ then$:$%
\begin{equation*}
m_{ii}^{t}(\mathbf{g}(S),\delta )=\left\{ 
\begin{array}{ccc}
1+m_{S}\delta ^{2}+k_{S}m_{S}^{2}\delta ^{4}+...+k_{S}^{\frac{t}{2}-1}m_{S}^{%
\frac{t}{2}}\delta ^{t}, & \text{if} & t\text{ is even}, \\ 
1+m_{S}\delta ^{2}+k_{S}m_{S}^{2}\delta ^{4}+...+k_{S}^{\frac{t-1}{2}%
-1}m_{S}^{\frac{t-1}{2}}\delta ^{t-1}, & \text{if} & t\text{ is odd and }t>1,%
\end{array}%
\right.
\end{equation*}%
\newline

Note that $m_{ii}^{0}(\mathbf{g}(S),\delta )=m_{ii}^{1}(\mathbf{g}(S),\delta
)=1.$

\item If $j\in M,$ then$:$ 
\begin{equation*}
m_{jj}^{t}(\mathbf{g}(S),\delta )=\left\{ 
\begin{array}{ccc}
1+k_{S}\delta ^{2}+k_{S}^{2}m_{S}\delta ^{4}+...+k_{S}^{\frac{t}{2}}m_{S}^{%
\frac{t}{2}-1}\delta ^{t}, & \text{if} & t\text{ is even}, \\ 
1+k_{S}\delta ^{2}+k_{S}^{2}m_{S}\delta ^{4}+...+k_{S}^{\frac{t-1}{2}}m_{S}^{%
\frac{t-1}{2}-1}\delta ^{t-1}, & \text{if} & t\text{ is odd and }t>1,%
\end{array}%
\right.
\end{equation*}

Note that $m_{jj}^{0}(\mathbf{g}(S),\delta )=m_{jj}^{1}(\mathbf{g}(S),\delta
)=1.$

\item If $i,j\in K$ and $i\neq j,$ then$:$ $\ $%
\begin{equation*}
m_{ij}^{t}(\mathbf{g}(S),\delta )=\left\{ 
\begin{array}{ccc}
m_{S}\delta ^{2}+k_{S}m_{S}^{2}\delta ^{4}+...+k_{S}^{\frac{t}{2}-1}m_{S}^{%
\frac{t}{2}}\delta ^{t}, & \text{if} & t\text{ is even}, \\ 
m_{S}\delta ^{2}+k_{S}m_{S}^{2}\delta ^{4}+...+k_{S}^{\frac{t-1}{2}-1}m_{S}^{%
\frac{t-1}{2}}\delta ^{t-1}, & \text{if} & t\text{ is odd and }t>1,%
\end{array}%
\right.
\end{equation*}

Note that $m_{ij}^{0}(\mathbf{g}(S),\delta )=m_{ij}^{1}(\mathbf{g}(S),\delta
)=0.$

\item If $i,j\in M$ and $i\neq j,$ then$:$ 
\begin{equation*}
m_{ij}^{t}(\mathbf{g}(S),\delta )=\left\{ 
\begin{array}{ccc}
k_{S}\delta ^{2}+k_{S}^{2}m_{S}\delta ^{4}+...+k_{S}^{\frac{t}{2}}m_{S}^{%
\frac{t}{2}-1}\delta ^{t}, & \text{if} & t\text{ is even}, \\ 
k_{S}\delta ^{2}+k_{S}^{2}m_{S}\delta ^{4}+...+k_{S}^{\frac{t-1}{2}}m_{S}^{%
\frac{t-1}{2}-1}\delta ^{t-1}, & \text{if} & t\text{ is odd and }t>1,%
\end{array}%
\right.
\end{equation*}

Note that $m_{ij}^{0}(\mathbf{g}(S),\delta )=m_{ij}^{1}(\mathbf{g}(S),\delta
)=0.$

\item If $i\in K$ and $j\in M$ \ or the opposite, then: 
\begin{equation*}
m_{ij}^{t}(\mathbf{g}(S),\delta )=\left\{ 
\begin{array}{ccc}
\delta +k_{S}m_{S}\delta ^{3}+...+k_{S}^{\frac{t}{2}-1}m_{S}^{\frac{t}{2}%
-1}\delta ^{t-1}, & \text{if} & t\text{ is even}, \\ 
\delta +k_{S}m_{S}\delta ^{3}+...+k_{S}^{\frac{t-1}{2}}m_{S}^{\frac{t-1}{2}%
}\delta ^{t}, & \text{if} & t\text{ is odd and }t>1,%
\end{array}%
\right.
\end{equation*}

Note that $m_{ij}^{0}(\mathbf{g}(S),\delta )=0,m_{ij}^{1}(\mathbf{g}%
(S),\delta )=\delta .$
\end{itemize}

Therefore, to calculate the productivity of the worker have $i\in S,$ we
need to consider four cases:

\begin{itemize}
\item[(1)] $t$ is even and $i\in K$
\end{itemize}

\begin{eqnarray*}
p_{i}^{S}(\delta ,t) &=&1+k_{S}\cdot \left( m_{S}\delta
^{2}+k_{S}m_{S}^{2}\delta ^{4}+...+k_{S}^{\frac{t}{2}-1}m_{S}^{\frac{t}{2}%
}\delta ^{t}\right) \\
&&+m_{S}\cdot \left( \delta +k_{S}m_{S}\delta ^{3}+...+k_{S}^{\frac{t}{2}%
-1}m_{S}^{\frac{t}{2}-1}\delta ^{t-1}\right) \\
&=&1+\left( k_{S}m_{S}\delta ^{2}+k_{S}^{2}m_{S}^{2}\delta ^{4}+...+k_{S}^{%
\frac{t}{2}}m_{S}^{\frac{t}{2}}\delta ^{t}\right) \\
&&+\left( m_{S}\delta +k_{S}m_{S}^{2}\delta ^{3}+...+k_{S}^{\frac{t}{2}%
-1}m_{S}^{\frac{t}{2}}\delta ^{t-1}\right) \\
&=&1+\overset{\frac{t}{2}}{\underset{u=1}{\sum }}k_{S}^{u}m_{S}^{u}\delta
^{2u}+\overset{\frac{t}{2}}{\underset{u=1}{\sum }}k_{S}^{u-1}m_{S}^{u}\delta
^{2u-1}
\end{eqnarray*}

\begin{itemize}
\item[(2)] $t$ is even and $i\in M$
\end{itemize}

\begin{eqnarray*}
p_{i}^{S}(\delta ,t) &=&1+m_{S}\cdot \left( k_{S}\delta
^{2}+k_{S}^{2}m_{S}\delta ^{4}+...+k_{S}^{\frac{t}{2}}m_{S}^{\frac{t}{2}%
-1}\delta ^{t}\right) \\
&&+k_{S}\cdot \left( \delta +k_{S}m_{S}\delta ^{3}+...+k_{S}^{\frac{t-2}{2}%
}m_{S}^{\frac{t-2}{2}}\delta ^{t-1}\right) \\
&=&1+\left( k_{S}m_{S}\delta ^{2}+k_{S}^{2}m_{S}^{2}\delta ^{4}+...+k_{S}^{%
\frac{t}{2}}m_{S}^{\frac{t}{2}}\delta ^{t}\right) \\
&&+\left( k_{S}\delta +k_{S}^{2}m_{S}\delta ^{3}+...+k_{S}^{\frac{t}{2}%
}m_{S}^{\frac{t}{2}-1}\delta ^{t-1}\right) \\
&=&1+\overset{\frac{t}{2}}{\underset{u=1}{\sum }}k_{S}^{u}m_{S}^{u}\delta
^{2u}+\overset{\frac{t}{2}}{\underset{u=1}{\sum }}k_{S}^{u}m_{S}^{u-1}\delta
^{2u-1}
\end{eqnarray*}

\begin{itemize}
\item[(3)] $t$ is odd and $i\in K$
\end{itemize}

\begin{eqnarray*}
p_{i}^{S}(\delta ,t) &=&1+k_{S}\cdot \left( m_{S}\delta
^{2}+k_{S}m_{S}^{2}\delta ^{4}+...+k_{S}^{\frac{t-1}{2}-1}m_{S}^{\frac{t-1}{2%
}}\delta ^{t-1}\right) \\
&&+m_{S}\cdot \left( \delta +k_{S}m_{S}\delta ^{3}+...+k_{S}^{\frac{t-1}{2}%
}m_{S}^{\frac{t-1}{2}}\delta ^{t}\right) \\
&=&1+\left( k_{S}m_{S}\delta ^{2}+k_{S}^{2}m_{S}^{2}\delta ^{4}+...+k_{S}^{%
\frac{t-1}{2}}m_{S}^{\frac{t-1}{2}}\delta ^{t-1}\right) \\
&&+\left( m_{S}\delta +k_{S}m_{S}^{2}\delta ^{3}+...+k_{S}^{\frac{t+1}{2}%
-1}m_{S}^{\frac{t+1}{2}}\delta ^{t}\right) \\
&=&1+\overset{\frac{t-1}{2}}{\underset{u=1}{\sum }}k_{S}^{u}m_{S}^{u}\delta
^{2u}+\overset{\frac{t+1}{2}}{\underset{u=1}{\sum }}k_{S}^{u-1}m_{S}^{u}%
\delta ^{2u-1}
\end{eqnarray*}

\begin{itemize}
\item[(4)] $t$ is odd and $i\in M$
\end{itemize}

\begin{eqnarray*}
p_{i}^{S}(\delta ,t) &=&1+m_{S}\cdot \left( k_{S}\delta
^{2}+k_{S}^{2}m_{S}\delta ^{4}+...+k_{S}^{\frac{t-1}{2}}m_{S}^{\frac{t-1}{2}%
-1}\delta ^{t-1}\right) \\
&&+k_{S}\cdot \left( \delta +k_{S}m_{S}\delta ^{3}+...+k_{S}^{\frac{t-1}{2}%
}m_{S}^{\frac{t-1}{2}}\delta ^{t}\right) \\
&=&1+\left( k_{S}m_{S}\delta ^{2}+k_{S}^{2}m_{S}^{2}\delta ^{4}+...+k_{S}^{%
\frac{t-1}{2}}m_{S}^{\frac{t-1}{2}}\delta ^{t-1}\right) \\
&&+\left( k_{S}\delta +k_{S}^{2}m_{S}\delta ^{3}+...+k_{S}^{\frac{t+1}{2}%
}m_{S}^{\frac{t+1}{2}-1}\delta ^{t}\right) \\
&=&1+\overset{\frac{t-1}{2}}{\underset{u=1}{\sum }}k_{S}^{u}m_{S}^{u}\delta
^{2u}+\overset{\frac{t+1}{2}}{\underset{u=1}{\sum }}k_{S}^{u}m_{S}^{u-1}%
\delta ^{2u-1}
\end{eqnarray*}

From the above results we can find an explicit form for the game $\left(N,v_{\delta }^{t}\right)$. Take $S \in N$, two cases are distinguished:

\begin{itemize}
\item If $t=0.$ It is straightforward by definition.

\item If $t>0$ is even%
\begin{eqnarray*}
v_{\delta }^{t}(S) &=&\sum_{i\in S}p_{i}^{S}(\delta ,t)=\sum\limits_{i\in
K(S)}p_{i}^{S}(\delta ,t)+\sum\limits_{i\in M(S)}p_{i}^{S}(\delta ,t) \\
&=&k_{S}\cdot \left( 1+\overset{\frac{t}{2}}{\underset{u=1}{\sum }}%
k_{S}^{u}m_{S}^{u}\delta ^{2u}+\overset{\frac{t}{2}}{\underset{u=1}{\sum }}%
k_{S}^{u-1}m_{S}^{u}\delta ^{2u-1}\right) \\
&&+m_{S}\cdot \left( 1+\overset{\frac{t}{2}}{\underset{u=1}{\sum }}%
k_{S}^{u}m_{S}^{u}\delta ^{2u}+\overset{\frac{t}{2}}{\underset{u=1}{\sum }}%
k_{S}^{u}m_{S}^{u-1}\delta ^{2u-1}\right) \\
&=&k_{S}+m_{S}+\left( k_{S}+m_{S}\right) \overset{\frac{t}{2}}{\underset{u=1}%
{\sum }}k_{S}^{u}m_{S}^{u}\delta ^{2u} \\
&&+\overset{\frac{t}{2}}{\underset{u=1}{\sum }}k_{S}^{u}m_{S}^{u}\delta
^{2u-1}+\overset{\frac{t}{2}}{\underset{u=1}{\sum }}k_{S}^{u}m_{S}^{u}\delta
^{2u-1} \\
&=&\left\vert S\right\vert +\left\vert S\right\vert \overset{\frac{t}{2}}{%
\underset{u=1}{\sum }}k_{S}^{u}m_{S}^{u}\delta ^{2u}+2\overset{\frac{t}{2}}{%
\underset{u=1}{\sum }}k_{S}^{u}m_{S}^{u}\delta ^{2u-1} \\
&=&\left\vert S\right\vert +\left( \left\vert S\right\vert \delta +2\right) 
\overset{\frac{t}{2}}{\underset{u=1}{\sum }}k_{S}^{u}m_{S}^{u}\delta ^{2u-1}
\end{eqnarray*}

\item If $t$ is odd: 
\begin{eqnarray*}
v_{\delta }^{t}(S) &=&\sum_{i\in S}p_{i}^{S}(\delta ,t)=\sum\limits_{i\in
K(S)}p_{i}^{S}(\delta ,t)+\sum\limits_{i\in M(S)}p_{i}^{S}(\delta ,t) \\
&=&k_{S}\cdot \left( 1+\overset{\frac{t-1}{2}}{\underset{u=1}{\sum }}%
k_{S}^{u}m_{S}^{u}\delta ^{2u}+\overset{\frac{t+1}{2}}{\underset{u=1}{\sum }}%
k_{S}^{u-1}m_{S}^{u}\delta ^{2u-1}\right) \\
&&+m_{S}\cdot \left( 1+\overset{\frac{t-1}{2}}{\underset{u=1}{\sum }}%
k_{S}^{u}m_{S}^{u}\delta ^{2u}+\overset{\frac{t+1}{2}}{\underset{u=1}{\sum }}%
k_{S}^{u}m_{S}^{u-1}\delta ^{2u-1}\right) \\
&=&k_{S}+m_{S}+\left( k_{S}+m_{S}\right) \overset{\frac{t-1}{2}}{\underset{%
u=1}{\sum }}k_{S}^{u}m_{S}^{u}\delta ^{2u} \\
&&+\overset{\frac{t+1}{2}}{\underset{u=1}{\sum }}k_{S}^{u}m_{S}^{u}\delta
^{2u-1}+\overset{\frac{t+1}{2}}{\underset{u=1}{\sum }}k_{S}^{u}m_{S}^{u}%
\delta ^{2u-1} \\
&=&\left\vert S\right\vert +\left\vert S\right\vert \overset{\frac{t-1}{2}}{%
\underset{u=1}{\sum }}k_{S}^{u}m_{S}^{u}\delta ^{2u}+2\overset{\frac{t+1}{2}}%
{\underset{u=1}{\sum }}k_{S}^{u}m_{S}^{u}\delta ^{2u-1} \\
&=&\left\vert S\right\vert +\left( \left\vert S\right\vert \delta +2\right) 
\overset{\frac{t-1}{2}}{\underset{u=1}{\sum }}k_{S}^{u}m_{S}^{u}\delta
^{2u-1}+2k_{S}^{\frac{t+1}{2}}m_{S}^{\frac{t+1}{2}}\delta ^{t}
\end{eqnarray*}
\end{itemize}

\hfill
\end{proof}

\bigskip

\noindent \begin{proof}[Proof of Theorem 3.2]
Consider $g=(K,M,E)$ a complete  bipartite network and its corresponding FAN
game $(N,v_{\delta }^{t})$. Take $S,T\subseteq N$ such that $S\subseteq T$
with $i\in S,$ then $k_{S}\leq k_{T}$ and $m_{S}\leq m_{T}.$ We have to
prove that $v_{\delta }^{t}(S)-v_{\delta }^{t}(S\backslash \{i\})\leq
v_{\delta }^{t}(T)-v_{\delta }^{t}(T\backslash \{i\}).$ Two cases are
distinguished:

\begin{itemize}
\item $t\ $is even. 
\begin{eqnarray*}
&&v_{\delta }^{t}(S)-v_{\delta }^{t}(S\backslash \{i\}) \\
&=&\left( \left\vert S\right\vert +\left( \left\vert S\right\vert \delta
+2\right) \overset{\frac{t}{2}}{\underset{u=1}{\sum }}k_{S}^{u}m_{S}^{u}%
\delta ^{2u-1}\right) \\
&&-\left( \left\vert S\backslash \{i\}\right\vert +\left( \left\vert
S\backslash \{i\}\right\vert \delta +2\right) \overset{\frac{t}{2}}{\underset%
{u=1}{\sum }}\left( k_{S}-1\right) ^{u}m_{S}^{u}\delta ^{2u-1}\right) \\
&=&1+\overset{\frac{t}{2}}{\underset{u=1}{\sum }}\left[ \left( \left(
\left\vert S\right\vert \delta +2\right) k_{S}^{u}-\left( \left\vert
S\right\vert \delta -\delta +2\right) \left( k_{S}-1\right) ^{u}\right)
m_{S}^{u}\delta ^{2u-1}\right] \\
&\leq &1+\overset{\frac{t}{2}}{\underset{u=1}{\sum }}\left[ \left( \left(
\left\vert S\right\vert \delta +2\right) k_{S}^{u}-\left( \left\vert
S\right\vert \delta -\delta +2\right) \left( k_{S}-1\right) ^{u}\right)
m_{T}^{u}\delta ^{2u-1}\right] \\
&\leq &\overset{\frac{t}{2}}{\underset{u=1}{\sum }}\left[ \left( \left(
\left\vert S\right\vert \delta +2\right) k_{T}^{u}-\left( \left\vert
S\right\vert \delta -\delta +2\right) \left( k_{T}-1\right) ^{u}\right)
m_{T}^{u}\delta ^{2u-1}\right] \\
&\leq &\overset{\frac{t}{2}}{\underset{u=1}{\sum }}\left[ \left( \left(
\left\vert T\right\vert \delta +2\right) k_{T}^{u}-\left( \left\vert
T\right\vert \delta -\delta +2\right) \left( k_{T}-1\right) ^{u}\right)
m_{T}^{u}\delta ^{2u-1}\right] =v_{\delta }^{t}(T)-v_{\delta
}^{t}(T\backslash \{i\})
\end{eqnarray*}

\item $t\ $is odd. A similar argument demonstrates it.
\end{itemize}

\hfill
\end{proof}

\bigskip

\noindent \begin{proof}[Proof of Lemma 4.1]
Consider $g=(K,M,E)$ a complete  bipartite network and $\Lambda (g,\delta )$
the set of all possible FAN games with index $\delta \geq 0$. For each $%
S\subseteq N,g(S)=(K(S),M(S),E(S))$ is a subnetwork of $g$. We distinguish
two cases.

If $\delta >0,$ then

\begin{eqnarray*}
\underset{t\rightarrow \infty }{\lim }v_{\delta }^{t}(S) &=&\left\vert
S\right\vert +\left\vert S\right\vert \overset{\infty }{\underset{u=1}{\sum }%
}\left( k_{S}m_{S}\delta ^{2}\right) ^{u}+\frac{2}{\delta }\overset{\infty }{%
\underset{u=1}{\sum }}\left( k_{S}m_{S}\delta ^{2}\right) ^{u} \\
&=&\left\vert S\right\vert +\left( \left\vert S\right\vert +\frac{2}{\delta }%
\right) \overset{\infty }{\underset{u=1}{\sum }}\left( k_{S}m_{S}\delta
^{2}\right) ^{u}
\end{eqnarray*}

$\overset{\infty }{\underset{u=1}{\sum }}\left( k_{S}m_{S}\delta ^{2}\right)
^{u}$ converges to $\frac{k_{S}m_{S}\delta ^{2}}{1-k_{S}m_{S}\delta ^{2}}$
if and only if $k_{S}m_{S}\delta ^{2}<1\Leftrightarrow \delta <\frac{1}{%
\sqrt{k_{S}m_{S}}}=\frac{1}{\lambda _{\max }(S)}.$ Hence,

\begin{eqnarray*}
\underset{t\rightarrow \infty }{\lim }v_{\delta }^{t}(S) &=&\left\vert
S\right\vert +\left( \left\vert S\right\vert +\frac{2}{\delta }\right) \frac{%
k_{S}m_{S}\delta ^{2}}{1-k_{S}m_{S}\delta ^{2}} \\
&=&k_{S}+m_{S}+\frac{\left( k_{S}+m_{S}\right) k_{S}m_{S}\delta ^{2}}{%
1-k_{S}m_{S}\delta ^{2}}+\frac{2k_{S}m_{S}\delta }{1-k_{S}m_{S}\delta ^{2}}
\\
&=&\frac{k_{S}+m_{S}-k_{S}^{2}m_{S}\delta ^{2}-k_{S}m_{S}\delta
^{2}+k_{S}^{2}m_{S}\delta ^{2}+k_{S}m_{S}^{2}\delta ^{2}+2k_{S}m_{S}\delta }{%
1-k_{S}m_{S}\delta ^{2}} \\
&=&\frac{k_{S}+m_{S}+2k_{S}m_{S}\delta }{1-k_{S}m_{S}\delta ^{2}}:=v_{\delta
}(S).
\end{eqnarray*}

If $\delta =0,$ then it is easy to check that $v_{\delta }^{t}(S)=\left\vert
S\right\vert :=v_{\delta }(S)$ for all $t\in \mathbb{N}$ and for each
coalition $S\subseteq N.$ \hfill
\end{proof}

\bigskip

\noindent \begin{proof}[Proof of Theorem 4.2]
Consider $g=(K,M,E)$ a complete  bipartite network and $(N,v_{\delta }^{t})$
his corresponding FAN game. For each $S\subseteq N,g(S)=(K(S),M(S),E(S))$ is
a subnetwork of $g$. We know that $\lambda _{\max }(N)\geq \lambda _{\max
}(S)$ for all $S\subseteq N.$ Hence, if $\delta \in \left[ 0,\frac{1}{%
\lambda _{\max }(N)}\right[ $, then $\delta \in \left[ 0,\frac{1}{\lambda
_{\max }(S)}\right[ $ for all $S\subseteq N,$ and so by Lemma \ref{lemma 1}
we conclude that $\left\{ v_{\delta }^{t}\right\} _{t\in \mathbb{N}}$
converges to $v_{\delta },$ defined as $v_{\delta }(S)=\frac{%
k_{S}+m_{S}+2k_{S}m_{S}\delta }{1-k_{S}m_{S}\delta ^{2}}$, for any $%
S\subseteq N.$ \hfill
\end{proof}

\bigskip

\noindent \begin{proof}[Proof of Proposition 4.4]
Take a coalition $S\subseteq N$. Then, $0\leq \delta <\frac{1}{\lambda
_{\max }(N)}\leq \frac{1}{\lambda _{\max }(S)}$. By the proof of Lemma \ref%
{lemma 1}, we know that $\overset{\infty }{\underset{u=1}{\sum }}\left(
k_{S}m_{S}\delta ^{2}\right) ^{u}$ converges to $\frac{k_{S}m_{S}\delta ^{2}%
}{1-k_{S}m_{S}\delta ^{2}}.$ Hence, if $i\in K(S),$ then 
\begin{eqnarray*}
\underset{t\rightarrow \infty }{\lim }p_{i}^{S}(\delta ,t) &=&1+\overset{%
\infty }{\underset{u=1}{\sum }}k_{S}^{u}m_{S}^{u}\delta ^{2u}+\overset{%
\infty }{\underset{u=1}{\sum }}k_{S}^{u-1}m_{S}^{u}\delta ^{2u-1} \\
&=&1+\left( k_{S}+\frac{1}{\delta }\right) \overset{\infty }{\underset{u=1}{%
\sum }}k_{S}^{u-1}m_{S}^{u}\delta ^{2u} \\
&=&1+\left( k_{S}+\frac{1}{\delta }\right) \frac{1}{k_{S}}\overset{\infty }{%
\underset{u=1}{\sum }}\left( k_{S}m_{S}\delta ^{2}\right) ^{u} \\
&=&1+\left( k_{S}+\frac{1}{\delta }\right) \frac{1}{k_{S}}\frac{%
k_{S}m_{S}\delta ^{2}}{1-k_{S}m_{S}\delta ^{2}} \\
&=&1+\frac{k_{S}m_{S}\delta ^{2}}{1-k_{S}m_{S}\delta ^{2}}+\frac{m_{S}\delta 
}{1-k_{S}m_{S}\delta ^{2}} \\
&=&\frac{1-k_{S}m_{S}\delta ^{2}+k_{S}m_{S}\delta ^{2}+m_{S}\delta }{%
1-k_{S}m_{S}\delta ^{2}}=\frac{1+m_{S}\delta }{1-k_{S}m_{S}\delta ^{2}}%
=:p_{i}^{S}(\delta ).
\end{eqnarray*}

For $j\in M(S),$ a similar argument proves that $\underset{t\rightarrow
\infty }{\lim }p_{j}^{S}(\delta ,t)=\frac{1+k_{S}\delta }{1-k_{S}m_{S}\delta
^{2}}=:p_{j}^{S}(\delta )$.

It is easy to prove that $p_{i}^{N}(\delta )\geq p_{i}^{S}(\delta )$ for all 
$S\subseteq N$. Indeed, if $i\notin S$ then $p_{i}^{S}(\delta )=0,$ and the
inequelity holds. If $i\notin S,$ the inequality is satisfied because $%
k_{N}\geq k_{S}$ and $m_{N}\geq m_{S}$. Therefore, if we take a coalition $%
S\subseteq N$, it is satisfies that: 
\begin{eqnarray*}
\sum_{i\in S}p_{i}^{N}(\delta ) &\geq &\sum_{i\in S}p_{i}^{S}(\delta
)=\sum_{i\in K(S)}p_{i}^{S}(\delta )+\sum_{i\in M(S)}p_{i}^{S}(\delta ) \\
&=&k_{S}\cdot \frac{1+m_{S}\delta }{1-k_{S}m_{S}\delta ^{2}}+m_{S}\cdot 
\frac{1+k_{S}\delta }{1-k_{S}m_{S}\delta ^{2}}=v_{\delta }(S).
\end{eqnarray*}%
It is straitgforward to prove that $\sum_{i\in N}p_{i}^{N}(\delta
)=v_{\delta }(N).$ Hence, $p^{N}(\delta )\in Core(N,v_{\delta })$.

If $\delta =0$ then, $v_{\delta }(S)=\left\vert S\right\vert $ for all team $%
S\subseteq N$ and $p_{i}^{N}(0)=1$ for each worker $i\in N.$ Therefore, $%
p^{N}(0)\in Core(N,v_{\delta }).$ We then conclude that $p^{N}(\delta )\in
Core(N,v_{\delta }),$ for any $\delta \geq 0.$ \hfill
\end{proof}

\bigskip

\noindent \begin{proof}[Proof of Lemma 4.5]
If $i\in K(S),$ then $v_{\delta }(S)-v_{\delta }(S\backslash \{i\})$ is
equal to:

\begin{align*}
& \frac{k_{S}+m_{S}+2k_{S}m_{S}\delta }{1-k_{S}m_{S}\delta ^{2}}-\frac{%
\left( k_{S}-1\right) +m_{S}+2\left( k_{S}-1\right) m_{S}\delta }{1-\left(
k_{S}-1\right) m_{S}\delta ^{2}} \\
& =\frac{k_{S}+m_{S}+2k_{S}m_{S}\delta }{1-k_{S}m_{S}\delta ^{2}}-\frac{%
k_{S}+m_{S}+2k_{S}m_{S}\delta -2m_{S}\delta -1}{1-k_{S}m_{S}\delta
^{2}+m_{S}\delta ^{2}} \\
& =\frac{P}{Q}-\frac{P}{Q+m_{S}\delta ^{2}}+\frac{2m_{S}\delta +1}{%
Q+m_{S}\delta ^{2}} \\
& =\frac{P(Q+m_{S}\delta ^{2})-PQ}{Q(Q+m_{S}\delta ^{2})}+\frac{2m_{S}\delta
+1}{Q+m_{S}\delta ^{2}}=\frac{P\cdot m_{S}\delta ^{2}}{Q(Q+m_{S}\delta ^{2})}%
+\frac{2m_{S}\delta +1}{Q+m_{S}\delta ^{2}} \\
& =\frac{k_{S}m_{S}\delta ^{2}+m_{S}^{2}\delta ^{2}+2k_{S}m_{S}^{2}\delta
^{3}+2m_{S}\delta +1-2k_{S}m_{S}^{2}\delta ^{3}-k_{S}m_{S}\delta ^{2}}{%
(1-k_{S}m_{S}\delta ^{2})(1-k_{S}m_{S}\delta ^{2}+m_{S}\delta ^{2})} \\
& =\frac{1+2m_{S}\delta +m_{S}^{2}\delta ^{2}}{(1-k_{S}m_{S}\delta
^{2})(1-k_{S}m_{S}\delta ^{2}+m_{S}\delta ^{2})}=\frac{\left( 1+m_{S}\delta
\right) ^{2}}{(1-k_{S}m_{S}\delta ^{2})(1-k_{S}m_{S}\delta ^{2}+m_{S}\delta
^{2})}
\end{align*}

where $P:=k_{S}+m_{S}+2k_{S}m_{S}\delta $ and $Q:=1-k_{S}m_{S}\delta ^{2}.$

If $i\in M(S),$ then $v_{\delta }(S)-v_{\delta }(S\backslash \{i\})$ is
equal to:

\begin{eqnarray*}
&&\frac{k_{S}+m_{S}+2k_{S}m_{S}\delta }{1-k_{S}m_{S}\delta ^{2}}-\frac{%
k_{S}+m_{S}-1+2k_{S}\left( m_{S}-1\right) \delta }{1-k_{S}\left(
m_{S}-1\right) \delta ^{2}} \\
&=&\frac{k_{S}+m_{S}+2k_{S}m_{S}\delta }{1-k_{S}m_{S}\delta ^{2}}-\frac{%
k_{S}+m_{S}+2k_{S}m_{S}\delta -2k_{S}\delta -1}{1-k_{S}m_{S}\delta
^{2}+k_{S}\delta ^{2}} \\
&=&\frac{P}{Q}-\frac{P}{Q+k_{S}\delta ^{2}}+\frac{2k_{S}\delta +1}{%
Q+k_{S}\delta ^{2}} \\
&=&\frac{P(Q+k_{S}\delta ^{2})-PQ}{Q(Q+k_{S}\delta ^{2})}+\frac{2k_{S}\delta
+1}{Q+k_{S}\delta ^{2}}=\frac{P\cdot k_{S}\delta ^{2}}{Q(Q+k_{S}\delta ^{2})}%
+\frac{2k_{S}\delta +1}{Q+k_{S}\delta ^{2}} \\
&=&\frac{k_{S}^{2}\delta ^{2}+k_{S}m_{S}\delta ^{2}+2k_{S}^{2}m_{S}\delta
^{3}+2k_{S}\delta +1-2k_{S}^{2}m_{S}\delta ^{3}-k_{S}m_{S}\delta ^{2}}{%
(1-k_{S}m_{S}\delta ^{2})(1-k_{S}m_{S}\delta ^{2}+k_{S}\delta ^{2})} \\
&=&\frac{k_{S}^{2}\delta ^{2}+2k_{S}\delta +1}{(1-k_{S}m_{S}\delta
^{2})(1-k_{S}m_{S}\delta ^{2}+k_{S}\delta ^{2})}=\frac{\left( 1+k_{S}\delta
\right) ^{2}}{(1-k_{S}m_{S}\delta ^{2})(1-k_{S}m_{S}\delta ^{2}+k_{S}\delta
^{2})}
\end{eqnarray*}%
\hfill
\end{proof}

\bigskip

\noindent \begin{proof}[Proof of Theorem 4.6]
Consider the AN game $(N,v_{\delta }).$ Let's demonstrate that for all $i\in S\subseteq T\subseteq N,v_{\delta
}(T)-v_{\delta }(T\backslash \{i\})\geq v_{\delta }(S)-v_{\delta
}(S\backslash \{i\}).$

Indeed, take $i\in S\subseteq T\subseteq N.$ If $i\in K(S)$ 
\begin{eqnarray*}
v_{\delta }(T)-v_{\delta }(T\backslash \{i\}) &=&\frac{\left( 1+m_{T}\delta
\right) ^{2}}{(1-k_{T}m_{T}\delta ^{2})(1-k_{T}m_{T}\delta ^{2}+m_{T}\delta
^{2})} \\
&=&\frac{\left( 1+m_{T}\delta \right) ^{2}}{(1-k_{T}m_{T}\delta
^{2})(1-\left( k_{T}-1\right) m_{T}\delta ^{2})} \\
&\geq &\frac{\left( 1+m_{S}\delta \right) ^{2}}{(1-k_{S}m_{S}\delta
^{2})(1-\left( k_{S}-1\right) m_{S}\delta ^{2})} \\
&=&\frac{\left( 1+m_{S}\delta \right) ^{2}}{(1-k_{S}m_{S}\delta
^{2})(1-k_{S}m_{S}\delta ^{2}+m_{S}\delta ^{2})} \\
&=&v_{\delta }(S)-v_{\delta }(S\backslash \{i\})
\end{eqnarray*}

since $k_{T}\geq k_{S}$ and $m_{T}\geq m_{S}.$ For $i\in M(S)$ the proof is
similar.\hfill
\end{proof}

\bigskip

\noindent \begin{proof}[Proof of Theorem 4.7]
We have that for all coalitions $S,R\subseteq N$ such that $k_{S}=k_{R}$ and 
$m_{S}=m_{R}$ then $v_{\delta }(S)=v_{\delta }(R).$ Moreover we can consider 
$\left\vert S\right\vert =m_{S}+k_{S},$ therefore $\gamma (S)=\frac{\left(
s-1\right) !(n-s)!}{n!}=\frac{\left( k_{S}+m_{S}-1\right) !(n-k_{S}-m_{S})!}{%
n!}=\gamma (k_{S},m_{S})$. If $i\in K$:%
\begin{eqnarray*}
\phi _{i}(v_{\delta }) &=&\sum\limits_{i\in S\subseteq N}\gamma (S)\cdot
\left( v_{\delta }(S)-v_{\delta }(S\backslash \{i\})\right) \\
&=&\sum\limits_{m=0}^{\left\vert M\right\vert }\sum\limits_{\substack{ %
i\in S\subseteq N:  \\ k_{S}=1\cap m_{S}=m}}\gamma (S)\cdot \left( v_{\delta
}(S)-v_{\delta }(S\backslash \{i\})\right) +... \\
&&+\sum\limits_{m=0}^{\left\vert M\right\vert }\sum\limits_{\substack{ %
i\in S\subseteq N:  \\ k_{S}=\left\vert K\right\vert \cap m_{S}=m}}\gamma
(S)\cdot \left( v_{\delta }(S)-v_{\delta }(S\backslash \{i\})\right) \\
&=&\sum\limits_{k=1}^{\left\vert K\right\vert
}\sum\limits_{m=0}^{\left\vert M\right\vert }\sum\limits_{\substack{ i\in
S\subseteq N:  \\ k_{S}=k\cap m_{S}=m}}\gamma (S)\cdot \frac{\left(
1+m\delta \right) ^{2}}{(1-km\delta ^{2})(1-km\delta ^{2}+m\delta ^{2})} \\
&=&\sum\limits_{k=1}^{\left\vert K\right\vert
}\sum\limits_{m=0}^{\left\vert M\right\vert }\frac{\left( 1+m\delta \right)
^{2}}{(1-km\delta ^{2})(1-km\delta ^{2}+m\delta ^{2})}\sum\limits 
_{\substack{ i\in S\subseteq N:  \\ k_{S}=k\cap m_{S}=m}}\gamma (k_{S},m_{S})
\\
&=&\sum\limits_{k=1}^{\left\vert K\right\vert
}\sum\limits_{m=0}^{\left\vert M\right\vert }\frac{\left( 1+m\delta \right)
^{2}}{(1-km\delta ^{2})(1-km\delta ^{2}+m\delta ^{2})}\cdot \binom{%
\left\vert M\right\vert }{m}\cdot \binom{\left\vert K\right\vert -1}{k-1}%
\cdot \gamma (k,m) \\
&=&\sum\limits_{k=1}^{\left\vert K\right\vert
}\sum\limits_{m=0}^{\left\vert M\right\vert }\Pi ^{K}_{M}(k,m)\cdot \frac{%
\left( 1+m\delta \right) ^{2}}{(1-km\delta ^{2})(1-km\delta ^{2}+m\delta
^{2})}
\end{eqnarray*}

If $i\in M:$%
\begin{eqnarray*}
\phi _{i}(v_{\delta }) &=&\sum\limits_{i\in S\subseteq N}\gamma (S)\cdot
\left( v_{\delta }(S)-v_{\delta }(S\backslash \{i\})\right) \\
&=&\sum\limits_{k=0}^{\left\vert K\right\vert }\sum\limits_{\substack{ %
i\in S\subseteq N:  \\ k_{S}=k\cap m_{S}=1}}\gamma (S)\cdot \left( v_{\delta
}(S)-v_{\delta }(S\backslash \{i\})\right) +... \\
&&+\sum\limits_{k=0}^{\left\vert K\right\vert }\sum\limits_{\substack{ %
i\in S\subseteq N:  \\ k_{S}=k\cap m_{S}=\left\vert M\right\vert }}\gamma
(S)\cdot \left( v_{\delta }(S)-v_{\delta }(S\backslash \{i\})\right) \\
&=&\sum\limits_{k=0}^{\left\vert K\right\vert
}\sum\limits_{m=1}^{\left\vert M\right\vert }\sum\limits_{\substack{ i\in
S\subseteq N:  \\ k_{S}=k\cap m_{S}=m}}\gamma (S)\cdot \frac{\left(
1+k\delta \right) ^{2}}{(1-km\delta ^{2})(1-km\delta ^{2}+k\delta ^{2})} \\
&=&\sum\limits_{k=0}^{\left\vert K\right\vert
}\sum\limits_{m=1}^{\left\vert M\right\vert }\frac{\left( 1+k\delta \right)
^{2}}{(1-km\delta ^{2})(1-km\delta ^{2}+k\delta ^{2})}\sum\limits 
_{\substack{ i\in S\subseteq N:  \\ k_{S}=k\cap m_{S}=m}}\gamma (k_{S},m_{S})
\\
&=&\sum\limits_{k=0}^{\left\vert K\right\vert
}\sum\limits_{m=1}^{\left\vert M\right\vert }\frac{\left( 1+k\delta \right)
^{2}}{(1-km\delta ^{2})(1-km\delta ^{2}+k\delta ^{2})}\cdot \binom{%
\left\vert M\right\vert -1}{m-1}\cdot \binom{\left\vert K\right\vert }{k}%
\cdot \gamma (k,m) \\
&=&\sum\limits_{k=0}^{\left\vert K\right\vert
}\sum\limits_{m=1}^{\left\vert M\right\vert }\Pi ^{M}_{K}(m,k)\cdot \frac{%
\left( 1+k\delta \right) ^{2}}{(1-km\delta ^{2})(1-km\delta ^{2}+k\delta
^{2})}
\end{eqnarray*}
\hfill
\end{proof}

\bigskip

\noindent \begin{proof}[Proof of Proposition 5.1]
Consider $g=(K,M,E)$ a complete  bipartite network and $(N,v_{\delta }^{t})$, 
$(N,v_{\delta }^{t-1})$ its corresponding FAN games. We distinguish two
cases.

Case 1: $t\ $is even, then

\begin{eqnarray*}
&&d_{\delta }^{t}(S)=v_{\delta }^{t}(S)-v_{\delta }^{t-1}(S)=\left(
\left\vert S\right\vert +\left( \left\vert S\right\vert \delta +2\right) 
\overset{\frac{t}{2}}{\underset{u=1}{\sum }}k_{S}^{u}m_{S}^{u}\delta
^{2u-1}\right) \\
&&-\left( \left\vert S\right\vert +\left( \left\vert S\right\vert \delta
+2\right) \overset{\frac{t}{2}-1}{\underset{u=1}{\sum }}\left(
k_{S}^{u}m_{S}^{u}\delta ^{2u-1}\right) +2k_{S}^{\frac{t}{2}}m_{S}^{\frac{1}{%
2}}\delta ^{t-1}\right) \\
&=&\left( \left\vert S\right\vert \delta +2\right) k_{S}^{\frac{t}{2}}m_{S}^{%
\frac{t}{2}}\delta ^{t-1}-2k_{S}^{\frac{t}{2}}m_{S}^{\frac{1}{2}}\delta
^{t-1}=\left\vert S\right\vert k_{S}^{\frac{t}{2}}m_{S}^{\frac{t}{2}}\delta
^{t}
\end{eqnarray*}

Case 2: $t\ $is odd, then

\begin{eqnarray*}
&&d_{\delta }^{t}(S)=v_{\delta }^{t}(S)-v_{\delta }^{t-1}(S) \\
&=&\left( \left\vert S\right\vert +\left( \left\vert S\right\vert \delta
+2\right) \overset{\frac{t-1}{2}}{\underset{u=1}{\sum }}\left(
k_{S}^{u}m_{S}^{u}\delta ^{2u-1}\right) +2k_{S}^{\frac{t+1}{2}}m_{S}^{\frac{%
t+1}{2}}\delta ^{t}\right) \\
&&-\left( \left\vert S\right\vert +\left( \left\vert S\right\vert \delta
+2\right) \overset{\frac{t-1}{2}}{\underset{u=1}{\sum }}k_{S}^{u}m_{S}^{u}%
\delta ^{2u-1}\right) =2k_{S}^{\frac{t+1}{2}}m_{S}^{\frac{t+1}{2}}\delta ^{t}
\end{eqnarray*}

\bigskip this can be rewritten as\bigskip 
\begin{equation*}
{d_{\delta }^{t}(S)=\left\{ 
\begin{array}{ccc}
\frac{k_{S}+m_{S}}{2}\cdot 2\left( \lambda _{\max }(S)\delta \right) ^{t}, & 
\text{if} & t\text{ is even}, \\ 
&  &  \\ 
\sqrt{k_{S}m_{S}}\cdot 2\left( \lambda _{\max }(S)\delta \right) ^{t}, & 
\text{if} & t\text{ is odd},%
\end{array}%
\right. }
\end{equation*}

We wonder if we can express both expressions for even and odd $t$ in a
single algebraic expression that depends on the eigenvalues. If this were
possible, we should be able to write both expressions in the form:

\begin{eqnarray*}
\frac{k_{S}+m_{S}}{2}\cdot 2\left( \lambda _{\max }(S)\delta \right) ^{t} &=&%
\left[ A\left( \lambda _{\max }(S)\right) ^{t}+B\left( -\lambda _{\max
}(S)\right) ^{t}\right] \delta ^{t}; \\
A+B &=&k_{S}+m_{S}\text{ (1)}
\end{eqnarray*}

and%
\begin{eqnarray*}
\sqrt{k_{S}m_{S}}\cdot 2\left( \lambda _{\max }(S)\delta \right) ^{t} &=&%
\left[ A\left( \lambda _{\max }(S)\right) ^{t}+B\left( -\lambda _{\max
}(S)\right) ^{t}\right] \delta ^{t} \\
A-B &=&2\sqrt{k_{S}m_{S}}\text{ \ (2)}
\end{eqnarray*}%
Solving the system (1)-(2), we obtain:

\begin{eqnarray*}
d_{t}^{\delta }(S) &=&\left[ \left( \frac{k_{S}+m_{S}}{2}+\sqrt{k_{S}m_{S}}%
\right) \left( \lambda _{\max }(S)\right) ^{t}+\left( \frac{k_{S}+m_{S}}{2}-%
\sqrt{k_{S}m_{S}}\right) \left( -\lambda _{\max }(S)\right) ^{t}\right] \
\delta ^{t} \\
&=&\frac{1}{2}\left[ \left( k_{S}+m_{S}+2\sqrt{k_{S}m_{S}}\right) \left(
\lambda _{\max }(S)\right) ^{t}+\left( k_{S}+m_{S}-2\sqrt{k_{S}m_{S}}\right)
\left( -\lambda _{\max }(S)\right) ^{t}\right] \ \delta ^{t} \\
&=&\frac{1}{2}\left[ \left( \sqrt{k_{S}}+\sqrt{m_{S}}\right) ^{2}\left(
\lambda _{\max }(S)\right) ^{t}+\left( \sqrt{k_{S}}-\sqrt{m_{S}}\right)
^{2}\left( -\lambda _{\max }(S)\right) ^{t}\right] \ \delta ^{t}
\end{eqnarray*}%
\hfill
\end{proof}

\bigskip

\noindent \begin{proof}[Proof of Proposition 5.2]
Consider $g=(K,M,E)$ a complete  bipartite network and $(N,d_{\delta }^{t})$
its corresponding difference game. Let%
\'{}%
s prove that $\sum_{i\in N}x_{i}^{t}(\delta )=d_{\delta }^{t}(N)$ and $%
\sum_{i\in S}x_{i}^{t}(\delta )\geq d_{\delta }^{t}(S).$ It is easy to check
that $x^{t}(\delta )$ satisfy efficiency:%
\begin{eqnarray*}
&&\sum_{i\in N}x_{i}^{t}(\delta )=\sum_{i\in K}\left( \frac{d_{\delta
}^{t}(N)}{\left\vert N\right\vert }\cdot \frac{\left\vert M\right\vert }{%
\left\vert K\right\vert }\right) +\sum_{i\in M}\left( \frac{d_{\delta
}^{t}(N)}{\left\vert N\right\vert }\cdot \frac{\left\vert K\right\vert }{%
\left\vert M\right\vert }\right) \\
&=&d_{\delta }^{t}(N)\cdot \left( \frac{\left\vert M\right\vert }{\left\vert
N\right\vert }+\frac{\left\vert K\right\vert }{\left\vert N\right\vert }%
\right) =d_{\delta }^{t}(N)
\end{eqnarray*}

It is straightforward to check that productivity distribution $x^{t}(\delta
) $ has the following explicit formula:

If $t$ is even, then 
\begin{equation*}
x_{i}^{t}(\delta )=\left\{ 
\begin{array}{ccc}
\left\vert K\right\vert ^{\frac{t}{2}-1}\left\vert M\right\vert ^{\frac{t}{2}%
+1}\delta ^{t}, & \text{if} & i\in K \\ 
&  &  \\ 
\left\vert K\right\vert ^{\frac{t}{2}+1}\left\vert M\right\vert ^{\frac{t}{2}%
-1}\delta ^{t}, & \text{if} & i\in M%
\end{array}%
\right.
\end{equation*}

if $t\ $is odd, then

\begin{equation*}
x_{i}^{t}(\delta )=\left\{ 
\begin{array}{ccc}
\frac{2}{\left\vert N\right\vert }\left\vert K\right\vert ^{\frac{t-1}{2}%
}\left\vert M\right\vert ^{\frac{t+3}{2}}\delta ^{t}, & \text{if} & i\in K
\\ 
&  &  \\ 
\frac{2}{\left\vert N\right\vert }\left\vert K\right\vert ^{\frac{t+3}{2}%
}\left\vert M\right\vert ^{\frac{t-1}{2}}\delta ^{t}, & \text{if} & i\in M%
\end{array}%
\right.
\end{equation*}

In order to demonstrate coalitional stability for a coalition $S\subset N,$
we distinguish two cases.

Case 1: $t\ $is even, then

\begin{eqnarray*}
&&\sum_{i\in S}x_{i}^{t}(\delta )=k_{S}\cdot \left\vert K\right\vert ^{\frac{%
t}{2}-1}\left\vert M\right\vert ^{\frac{t}{2}+1}\delta ^{t}+m_{S}\cdot
\left\vert K\right\vert ^{\frac{t}{2}+1}\left\vert M\right\vert ^{\frac{t}{2}%
-1}\delta ^{t} \\
&\geq &k_{S}\cdot k_{S}^{\frac{t}{2}-1}m_{S}^{\frac{t}{2}+1}\delta
^{t}+m_{S}\cdot k_{S}^{\frac{t}{2}+1}m_{S}^{\frac{t}{2}-1}\delta ^{t} \\
&=&k_{S}^{\frac{t}{2}}m_{S}^{\frac{t}{2}+1}\delta ^{t}+k_{S}^{\frac{t}{2}%
+1}m_{S}^{\frac{t}{2}}\delta ^{t}=\left\vert S\right\vert k_{S}^{\frac{t}{2}%
}m_{S}^{\frac{t}{2}}\delta ^{t}=d_{\delta }^{t}(S)\text{ .}
\end{eqnarray*}

Case 2: $t\ $is odd, then

\begin{eqnarray*}
&&\sum_{i\in S}x_{i}^{t}(\delta )=k_{S}\cdot \frac{2}{\left\vert
N\right\vert }\left\vert K\right\vert ^{\frac{t-1}{2}}\left\vert
M\right\vert ^{\frac{t+3}{2}}\delta ^{t}+m_{S}\cdot \frac{2}{\left\vert
N\right\vert }\left\vert K\right\vert ^{\frac{t+3}{2}}\left\vert
M\right\vert ^{\frac{t-1}{2}}\delta ^{t} \\
&=&k_{S}\cdot \frac{2\left\vert M\right\vert }{\left\vert N\right\vert }%
\left\vert K\right\vert ^{\frac{t-1}{2}}\left\vert M\right\vert ^{\frac{t+1}{%
2}}\delta ^{t}+m_{S}\cdot \frac{2\left\vert K\right\vert }{\left\vert
N\right\vert }\left\vert K\right\vert ^{\frac{t+1}{2}}\left\vert
M\right\vert ^{\frac{t-1}{2}}\delta ^{t} \\
&\geq &k_{S}\cdot \frac{2\left\vert M\right\vert }{\left\vert N\right\vert }%
k_{S}^{\frac{t-1}{2}}m_{S}^{\frac{t+1}{2}}\delta ^{t}+m_{S}\cdot \frac{%
2\left\vert K\right\vert }{\left\vert N\right\vert }k_{S}^{\frac{t+1}{2}%
}m_{S}^{\frac{t-1}{2}}\delta ^{t} \\
&=&\frac{2\left\vert M\right\vert }{\left\vert N\right\vert }k_{S}^{\frac{t+1%
}{2}}m_{S}^{\frac{t+1}{2}}\delta ^{t}+\frac{2\left\vert K\right\vert }{%
\left\vert N\right\vert }k_{S}^{\frac{t+1}{2}}m_{S}^{\frac{t+1}{2}}\delta
^{t}=2k_{S}^{\frac{t+1}{2}}m_{S}^{\frac{t+1}{2}}\delta ^{t}=d_{\delta
}^{t}(S)\text{ .}
\end{eqnarray*}%
\hfill
\end{proof}

\bigskip

\noindent \begin{proof}[Proof of Proposition 5.4]
Consider $g=(K,M,E)$ a complete  bipartite network and $\Lambda (\delta )$
the set of all possible FAN games with index $\delta >0$. We distinguish two
cases.

Case 1: $i\in K$, then 
\begin{eqnarray*}
\omega _{i}(\delta ) &=&1+\underset{t\rightarrow \infty }{\lim }\left( 
\overset{\frac{t}{2}}{\underset{u=1}{\sum }}\left( \left\vert K\right\vert
^{u-1}\left\vert M\right\vert ^{u+1}\delta ^{2u}\right) +\overset{\frac{t-1}{%
2}}{\underset{u=0}{\sum }}\left( \frac{2}{\left\vert N\right\vert }%
\left\vert K\right\vert ^{u}\left\vert M\right\vert ^{u+2}\delta
^{2u+1}\right) \right) \\
&=&1+\frac{\left\vert M\right\vert }{\left\vert K\right\vert }\underset{%
t\rightarrow \infty }{\lim }\left( \overset{\frac{t}{2}}{\underset{u=1}{\sum 
}}\left( \left\vert K\right\vert \left\vert M\right\vert \delta ^{2}\right)
^{u}\right) +\frac{2\left\vert M\right\vert ^{2}\delta }{\left\vert
N\right\vert }\underset{t\rightarrow \infty }{\lim }\left( \overset{\frac{t-1%
}{2}}{\underset{u=0}{\sum }}\left( \left\vert K\right\vert \left\vert
M\right\vert \delta ^{2}\right) ^{u}\right) \\
&=&1+\frac{\left\vert M\right\vert }{\left\vert K\right\vert }\overset{%
\infty }{\underset{u=1}{\sum }}\left( \left\vert K\right\vert \left\vert
M\right\vert \delta ^{2}\right) ^{u}+\frac{2\left\vert M\right\vert
^{2}\delta }{\left\vert N\right\vert }\overset{\infty }{\underset{u=0}{\sum }%
}\left( \left\vert K\right\vert \left\vert M\right\vert \delta ^{2}\right)
^{u} \\
&=&1+\frac{\left\vert M\right\vert }{\left\vert K\right\vert }\frac{%
\left\vert K\right\vert \left\vert M\right\vert \delta ^{2}}{1-\left\vert
K\right\vert \left\vert M\right\vert \delta ^{2}}+\frac{2\left\vert
M\right\vert ^{2}\delta }{\left\vert N\right\vert }\frac{1}{1-\left\vert
K\right\vert \left\vert M\right\vert \delta ^{2}} \\
&=&1+\left( \frac{\left\vert M\right\vert }{\left\vert K\right\vert }\delta +%
\frac{2\left\vert M\right\vert }{\left\vert N\right\vert \left\vert
K\right\vert }\right) \frac{\left\vert K\right\vert \left\vert M\right\vert
\delta }{1-\left\vert K\right\vert \left\vert M\right\vert \delta ^{2}}
\end{eqnarray*}

Case 2: $i\in M,$ then

\begin{eqnarray*}
\omega _{i}(\delta ) &=&1+\underset{t\rightarrow \infty }{\lim }\left( 
\overset{\frac{t}{2}}{\underset{u=1}{\sum }}\left( \left\vert K\right\vert
^{u+1}\left\vert M\right\vert ^{u-1}\delta ^{2u}\right) +\overset{\frac{t-1}{%
2}}{\underset{u=0}{\sum }}\left( \frac{2}{\left\vert N\right\vert }%
\left\vert K\right\vert ^{u+2}\left\vert M\right\vert ^{u}\delta
^{2u+1}\right) \right) \\
&=&1+\frac{\left\vert K\right\vert }{\left\vert M\right\vert }\underset{%
t\rightarrow \infty }{\lim }\left( \overset{\frac{t}{2}}{\underset{u=1}{\sum 
}}\left( \left\vert K\right\vert \left\vert M\right\vert \delta ^{2}\right)
^{u}\right) +\frac{2\left\vert K\right\vert ^{2}\delta }{\left\vert
N\right\vert }\underset{t\rightarrow \infty }{\lim }\left( \overset{\frac{t-1%
}{2}}{\underset{u=0}{\sum }}\left( \left\vert K\right\vert \left\vert
M\right\vert \delta ^{2}\right) ^{u}\right) \\
&=&1+\frac{\left\vert K\right\vert }{\left\vert M\right\vert }\overset{%
\infty }{\underset{u=1}{\sum }}\left( \left\vert K\right\vert \left\vert
M\right\vert \delta ^{2}\right) ^{u}+\frac{2\left\vert K\right\vert
^{2}\delta }{\left\vert N\right\vert }\overset{\infty }{\underset{u=0}{\sum }%
}\left( \left\vert K\right\vert \left\vert M\right\vert \delta ^{2}\right)
^{u} \\
&=&1+\frac{\left\vert K\right\vert }{\left\vert M\right\vert }\frac{%
\left\vert K\right\vert \left\vert M\right\vert \delta ^{2}}{1-\left\vert
K\right\vert \left\vert M\right\vert \delta ^{2}}^{u}+\frac{2\left\vert
K\right\vert ^{2}\delta }{\left\vert N\right\vert }\frac{1}{1-\left\vert
K\right\vert \left\vert M\right\vert \delta ^{2}} \\
&=&1+\left( \frac{\left\vert K\right\vert }{\left\vert M\right\vert }\delta +%
\frac{2\left\vert K\right\vert }{\left\vert N\right\vert \left\vert
M\right\vert }\right) \frac{\left\vert K\right\vert \left\vert M\right\vert
\delta }{1-\left\vert K\right\vert \left\vert M\right\vert \delta ^{2}}
\end{eqnarray*}
\hfill
\end{proof}

\bigskip

\noindent \begin{proof}[Proof of Theorem 5.5]
Consider $g=(K,M,E)$ a complete  bipartite network and $(N,v_{\delta })$ its
corresponding AN game. We prove first that $\omega (\delta )$ satisfies
efficiency. Indeed,

\begin{eqnarray*}
\sum_{i\in N}\omega _{i}(\delta ) &=&\sum_{i\in K}\omega _{i}(\delta
)+\sum_{i\in M}\omega _{i}(\delta ) \\
&=&\left\vert K\right\vert \left[ 1+\left( \frac{\left\vert M\right\vert }{%
\left\vert K\right\vert }\delta +\frac{2\left\vert M\right\vert }{\left\vert
N\right\vert \left\vert K\right\vert }\right) \frac{\left\vert K\right\vert
\left\vert M\right\vert \delta }{1-\left\vert K\right\vert \left\vert
M\right\vert \delta ^{2}}\right] \\
&&+\left\vert M\right\vert \left[ 1+\left( \frac{\left\vert K\right\vert }{%
\left\vert M\right\vert }\delta +\frac{2\left\vert K\right\vert }{\left\vert
N\right\vert \left\vert M\right\vert }\right) \frac{\left\vert K\right\vert
\left\vert M\right\vert \delta }{1-\left\vert K\right\vert \left\vert
M\right\vert \delta ^{2}}\right] \\
&=&\left\vert K\right\vert +\left\vert M\right\vert +\left( \left\vert
K\right\vert \delta +\left\vert M\right\vert \delta +\frac{2\left\vert
M\right\vert }{\left\vert N\right\vert }+\frac{2\left\vert K\right\vert }{%
\left\vert N\right\vert }\right) \frac{\left\vert K\right\vert \left\vert
M\right\vert \delta }{1-\left\vert K\right\vert \left\vert M\right\vert
\delta ^{2}} \\
&=&\left\vert K\right\vert +\left\vert M\right\vert +\left( \left\vert
K\right\vert \delta +\left\vert M\right\vert \delta +2\right) \frac{%
\left\vert K\right\vert \left\vert M\right\vert \delta }{1-\left\vert
K\right\vert \left\vert M\right\vert \delta ^{2}} \\
&=&\frac{\left\vert K\right\vert -\left\vert K\right\vert ^{2}\left\vert
M\right\vert \delta ^{2}+\left\vert M\right\vert -\left\vert K\right\vert
\left\vert M\right\vert ^{2}\delta ^{2}+\left\vert K\right\vert
^{2}\left\vert M\right\vert \delta ^{2}+\left\vert K\right\vert \left\vert
M\right\vert ^{2}\delta ^{2}+2\left\vert K\right\vert \left\vert
M\right\vert \delta }{1-\left\vert K\right\vert \left\vert M\right\vert
\delta ^{2}} \\
&=&\frac{\left\vert K\right\vert +\left\vert M\right\vert +2\left\vert
K\right\vert \left\vert M\right\vert \delta }{1-\left\vert K\right\vert
\left\vert M\right\vert \delta ^{2}}=v_{\delta }(N)
\end{eqnarray*}

Consider now the set of all possible FAN games with index $\delta .$ We know
that $x^{t}(\delta )$ is a core allocation of the game $(N,d_{\delta }^{t})$
for all $t\geq 1$. Moreover, $\overset{t}{\underset{u=1}{\sum }}d_{\delta
}^{u}=v_{\delta }^{t}-v_{\delta }^{0}$ and $\overset{t}{\underset{u=1}{\sum }%
}Core(N,d_{\delta }^{u})\varsubsetneq Core(N,v_{\delta }^{t}-v_{\delta
}^{0}).$ Hence, 
\begin{equation*}
\sum_{i\in S}\overset{t}{\underset{u=1}{\sum }}x_{i}^{u}(\delta )\geq
v_{\delta }^{t}(S)-v_{\delta }^{0}(S).
\end{equation*}

Then we take as $t$ tends to infinity,

\begin{eqnarray*}
\sum_{i\in S}\left( \omega _{i}(\delta )-1\right) &\geq &v_{\delta
}(S)-v_{\delta }^{0}(S); \\
\left( \sum_{i\in S}\omega _{i}(\delta )\right) -\left\vert S\right\vert
&\geq &v_{\delta }(S)-\left\vert S\right\vert ; \\
\sum_{i\in S}\omega _{i}(\delta ) &\geq &v_{\delta }(S)
\end{eqnarray*}

Hence, we conclude that $\omega (\delta )$ is a core allocation of $%
(N,v_{\delta })$.\hfill
\end{proof}

\bigskip

\noindent \begin{proof}[Proof of Theorem 5.7]
It is clear that the LRP distibution $\omega (\delta )$ satisfies EF, EB and
LBP.

To show the converse, take a productivity doistribution $\varphi $ on the
class of AN games, that satisfies EF, EB and LBP.

By EF and EB we have that $\left\vert K\right\vert \varphi _{i}(v_{\delta
})+\left\vert M\right\vert \varphi _{j}(v_{\delta })=v_{\delta }(N)$ for any 
$i\in K$ and $j\in M.$

Moreover, by EB and LBP: $\frac{\left\vert K\right\vert }{\left\vert
M\right\vert }\left( \left\vert K\right\vert \varphi _{i}(v_{\delta
})-\left\vert K\right\vert \right) =\left\vert M\right\vert \varphi
_{j}(v_{\delta })-\left\vert M\right\vert $ for any $i\in K$ and $j\in M.$
Substituting the second equation into the first equation, we obtain that:

\begin{eqnarray*}
\left\vert K\right\vert \varphi _{i}(v_{\delta }) &=&v_{\delta
}(N)-\left\vert M\right\vert \varphi _{j}(v_{\delta }) \\
\left\vert K\right\vert \varphi _{i}(v_{\delta }) &=&v_{\delta }(N)-\frac{%
\left\vert K\right\vert }{\left\vert M\right\vert }\left( \left\vert
K\right\vert \varphi _{i}(v_{\delta })-\left\vert K\right\vert \right)
-\left\vert M\right\vert ; \\
\varphi _{i}(v_{\delta }) &=&\frac{1}{\left\vert K\right\vert }v_{\delta
}(N)-\frac{1}{\left\vert M\right\vert }\left( \left\vert K\right\vert
\varphi _{i}(v_{\delta })-\left\vert K\right\vert \right) -\frac{\left\vert
M\right\vert }{\left\vert K\right\vert }; \\
\varphi _{i}(v_{\delta })+\frac{\left\vert K\right\vert }{\left\vert
M\right\vert }\varphi _{i}(v_{\delta }) &=&\frac{1}{\left\vert K\right\vert }%
v_{\delta }(N)+\frac{\left\vert K\right\vert }{\left\vert M\right\vert }-%
\frac{\left\vert M\right\vert }{\left\vert K\right\vert }; \\
\varphi _{i}(v_{\delta }) &=&\left( \frac{1}{\left\vert K\right\vert }%
v_{\delta }(N)+\frac{\left\vert K\right\vert }{\left\vert M\right\vert }-%
\frac{\left\vert M\right\vert }{\left\vert K\right\vert }\right) :\left( 1+%
\frac{\left\vert K\right\vert }{\left\vert M\right\vert }\right) .
\end{eqnarray*}

Developing the last expression, we obtain:

\begin{eqnarray*}
\varphi _{i}(v_{\delta }) &=&\left( \frac{1}{\left\vert K\right\vert }%
v_{\delta }(N)+\frac{\left\vert K\right\vert }{\left\vert M\right\vert }-%
\frac{\left\vert M\right\vert }{\left\vert K\right\vert }\right) :\left( 1+%
\frac{\left\vert K\right\vert }{\left\vert M\right\vert }\right) \\
&=&\left( \frac{\left\vert K\right\vert +\left\vert M\right\vert
+2\left\vert K\right\vert \left\vert M\right\vert \delta }{\left(
1-\left\vert K\right\vert \left\vert M\right\vert \delta ^{2}\right)
\left\vert K\right\vert }+\frac{\left\vert K\right\vert ^{2}-\left\vert
M\right\vert ^{2}}{\left\vert M\right\vert \left\vert K\right\vert }\right) :%
\frac{\left\vert N\right\vert }{\left\vert M\right\vert } \\
&=&\left( \frac{\left\vert N\right\vert +2\left\vert K\right\vert \left\vert
M\right\vert \delta }{\left( 1-\left\vert K\right\vert \left\vert
M\right\vert \delta ^{2}\right) \left\vert K\right\vert }+\frac{\left\vert
N\right\vert \left( \left\vert K\right\vert -\left\vert M\right\vert \right) 
}{\left\vert M\right\vert \left\vert K\right\vert }\right) \cdot \frac{%
\left\vert M\right\vert }{\left\vert N\right\vert } \\
&&\frac{\left\vert M\right\vert \left\vert N\right\vert +2\left\vert
K\right\vert \left\vert M\right\vert ^{2}\delta }{\left( 1-\left\vert
K\right\vert \left\vert M\right\vert \delta ^{2}\right) \left\vert
K\right\vert \left\vert N\right\vert }+\frac{\left\vert K\right\vert
-\left\vert M\right\vert }{\left\vert K\right\vert } \\
&=&\frac{\left\vert M\right\vert \left\vert N\right\vert +2\left\vert
K\right\vert \left\vert M\right\vert ^{2}\delta +\left( \left\vert
K\right\vert -\left\vert M\right\vert \right) \cdot \left( 1-\left\vert
K\right\vert \left\vert M\right\vert \delta ^{2}\right) \left\vert
N\right\vert }{\left( 1-\left\vert K\right\vert \left\vert M\right\vert
\delta ^{2}\right) \left\vert K\right\vert \left\vert N\right\vert } \\
&=&\frac{\left\vert M\right\vert \left\vert N\right\vert +2\left\vert
K\right\vert \left\vert M\right\vert ^{2}\delta +\left\vert K\right\vert
\left\vert N\right\vert -\left\vert K\right\vert ^{2}\left\vert M\right\vert
\delta ^{2}\left\vert N\right\vert -\left\vert M\right\vert \left\vert
N\right\vert +\left\vert K\right\vert \left\vert M\right\vert ^{2}\delta
^{2}\left\vert N\right\vert }{\left( 1-\left\vert K\right\vert \left\vert
M\right\vert \delta ^{2}\right) \left\vert K\right\vert \left\vert
N\right\vert } \\
&=&\frac{2\left\vert K\right\vert \left\vert M\right\vert ^{2}\delta
+\left\vert K\right\vert \left\vert N\right\vert -\left\vert K\right\vert
^{2}\left\vert M\right\vert \delta ^{2}\left\vert N\right\vert +\left\vert
K\right\vert \left\vert M\right\vert ^{2}\delta ^{2}\left\vert N\right\vert 
}{\left( 1-\left\vert K\right\vert \left\vert M\right\vert \delta
^{2}\right) \left\vert K\right\vert \left\vert N\right\vert } \\
&=&\frac{\left\vert K\right\vert \left\vert N\right\vert -\left\vert
K\right\vert ^{2}\left\vert M\right\vert \delta ^{2}\left\vert N\right\vert 
}{\left( 1-\left\vert K\right\vert \left\vert M\right\vert \delta
^{2}\right) \left\vert K\right\vert \left\vert N\right\vert }+\frac{%
\left\vert K\right\vert \left\vert M\right\vert ^{2}\delta ^{2}\left\vert
N\right\vert }{\left( 1-\left\vert K\right\vert \left\vert M\right\vert
\delta ^{2}\right) \left\vert K\right\vert \left\vert N\right\vert }+\frac{%
2\left\vert K\right\vert \left\vert M\right\vert ^{2}\delta }{\left(
1-\left\vert K\right\vert \left\vert M\right\vert \delta ^{2}\right)
\left\vert K\right\vert \left\vert N\right\vert } \\
&=&\frac{\left( 1-\left\vert K\right\vert \left\vert M\right\vert \delta
^{2}\right) \left\vert N\right\vert \left\vert K\right\vert }{\left(
1-\left\vert K\right\vert \left\vert M\right\vert \delta ^{2}\right)
\left\vert N\right\vert \left\vert K\right\vert }+\frac{\left\vert
K\right\vert \left\vert M\right\vert ^{2}\delta ^{2}}{\left( 1-\left\vert
K\right\vert \left\vert M\right\vert \delta ^{2}\right) \left\vert
K\right\vert }+\frac{2\left\vert K\right\vert \left\vert M\right\vert
^{2}\delta }{\left( 1-\left\vert K\right\vert \left\vert M\right\vert \delta
^{2}\right) \left\vert N\right\vert \left\vert K\right\vert } \\
&=&1+\left( \frac{\left\vert M\right\vert }{\left\vert K\right\vert }\delta +%
\frac{2\left\vert M\right\vert }{\left\vert N\right\vert \left\vert
K\right\vert }\right) \frac{\left\vert K\right\vert \left\vert M\right\vert
\delta }{1-\left\vert K\right\vert \left\vert M\right\vert \delta ^{2}}%
=\omega _{i}(\delta )
\end{eqnarray*}

Finally, 
\begin{eqnarray*}
\frac{\left\vert K\right\vert }{\left\vert M\right\vert }\left( \left\vert
K\right\vert \varphi _{i}(v_{\delta })-\left\vert K\right\vert \right)
&=&\left\vert M\right\vert \varphi _{j}(v_{\delta })-\left\vert M\right\vert
; \\
\varphi _{j}(v_{\delta }) &=&1+\frac{\left\vert K\right\vert ^{2}}{%
\left\vert M\right\vert ^{2}}\left( \varphi _{i}(v_{\delta })-1\right) ;
\end{eqnarray*}

Developing the expression, we obtain:%
\begin{eqnarray*}
\varphi _{j}(v_{\delta }) &=&1+\frac{\left\vert K\right\vert ^{2}}{%
\left\vert M\right\vert ^{2}}\left[ \left( \frac{\left\vert M\right\vert }{%
\left\vert K\right\vert }\delta +\frac{2\left\vert M\right\vert }{\left\vert
N\right\vert \left\vert K\right\vert }\right) \frac{\left\vert K\right\vert
\left\vert M\right\vert \delta }{1-\left\vert K\right\vert \left\vert
M\right\vert \delta ^{2}}\right] \\
&=&1+\left( \frac{\left\vert K\right\vert }{\left\vert M\right\vert }\delta +%
\frac{2\left\vert K\right\vert }{\left\vert N\right\vert \left\vert
M\right\vert }\right) \frac{\left\vert K\right\vert \left\vert M\right\vert
\delta }{1-\left\vert K\right\vert \left\vert M\right\vert \delta ^{2}}%
=\omega _{j}(\delta )
\end{eqnarray*}
\hfill
\end{proof}

\bigskip

\end{document}